\documentclass[conference]{IEEEtran}
\IEEEoverridecommandlockouts

\usepackage{cite}
\usepackage{amsmath,amssymb,amsfonts}
\usepackage{graphicx}
\usepackage{textcomp}
\usepackage{xcolor}
\usepackage{booktabs}
\usepackage{multirow}
\usepackage{subfig}
\usepackage{float}
\usepackage[printonlyused]{acronym}
\usepackage{url}
\usepackage{hyperref}

\hypersetup{colorlinks=true, linkcolor=blue, citecolor=blue, urlcolor=blue}

\begin{document}

\title{Catastrophic Learning: A New Attack Vector on Continual Learning Networks}

\author{\IEEEauthorblockN{Benedikt Kluss}
\IEEEauthorblockA{\textit{Chair for System Security} \\
\textit{University of Bonn}\\
Bonn, Germany \\
kluss@cs.uni-bonn.de}
\and
\IEEEauthorblockN{Niklas Bunzel}
\IEEEauthorblockA{\textit{Fraunhofer SIT / TU Darmstadt}\\
Darmstadt, Germany\\
bunzel@sit.fraunhofer.de}
}

\maketitle

\begin{abstract}
Continual Learning (CL) enables deep learning models to iteratively learn from a stream of data without forgetting prior knowledge. Existing adversarial research on CL primarily aims to re-enable catastrophic forgetting, thus attacking stability and reducing availability. In this work, we identify a novel security flaw: data manipulated by an attacker can reduce the learnability of the current or of upcoming iterations. We term such manipulations \emph{learning blockers}, as they specifically attack the plasticity of CL algorithms. Learning blockers are particularly harmful because they are difficult to detect during training and evaluation of the current iteration, since they can target iterations whose data the model has not yet encountered. When learning blockers additionally induce catastrophic forgetting, the resulting overall degradation is what we call \emph{catastrophic learning}. We formalize this scenario, define an associated threat model, and propose six attack strategies: Label-Exchange, Tensor-Exchange, Attraction-Coincident, Attraction-Preceding, Repulsion-Coincident, and Repulsion-Preceding. The Attraction variants minimize the loss between the poisoned and the victim iteration label, pulling their representations together in feature space; the Repulsion variants maximize this loss, pushing them apart so the stability mechanism resists the required parameter shift. In the Coincident variants, the poisoned and the victim iteration coincide, using a clean reference iteration only as a label source; in the Preceding variants, the poisoned iteration precedes the victim, leaving it clean but not learnable due to distorted representations learned beforehand. We evaluate the attacks on MNIST and Split-CIFAR10 against three CL strategies---DER, ER-ACE, and iCaRL---across more than 4{,}480 simulations. Our results demonstrate a strong vulnerability of CL algorithms: an adversary can selectively impede plasticity to hinder the acquisition of new knowledge, while simultaneously promoting the loss of previously learned information, thereby inducing a catastrophic learning scenario.
\end{abstract}


\section{Introduction}
\label{sec:intro}

In classical computer science, developers implement models to describe complex structures of the real world. With the rise of artificial intelligence, more and more of the world's complexity can be reproduced using data that goes far beyond the boundaries of classical deterministic algorithms. Thanks to machine learning, and especially deep learning, many research problems such as speech recognition, natural language processing, or computer vision can be solved~\cite{DBLP:journals/corr/abs-2108-00401}.

The acquisition of data is a crucial step in building a deep learning model with high accuracy. This is particularly true in critical sectors such as healthcare, autonomous driving, and cybersecurity, where the quality of data is essential to avoid incorrect decisions. Furthermore, data distributions are dynamic, as evidenced by the constant discovery of new research results. Consequently, a continual learning strategy is required in the field of deep learning. The concept of Continual Learning (CL), also referred to as incremental learning and lifelong learning~\cite{NEURIPS2022_20f44da8}, aims to address this challenge by enabling deep learning models to learn iteratively from a stream of data without forgetting prior knowledge. The objective is to facilitate seamless adaptation to novel scenarios without requiring extensive resources, such as memory or computing power, for comprehensive retraining. However, training a model on new data can lead to a phenomenon known as catastrophic forgetting~\cite{MCCLOSKEY1989109}, which causes misclassifications of previously acquired knowledge. Several strategies have been developed to mitigate this problem by preserving prior knowledge through the maintenance of model \emph{stability}, while simultaneously allowing the model to adapt to new information. This adaptation to new data is termed \emph{plasticity} in the scientific literature~\cite{8953627}.

In order to obtain new data, two factors must be considered: the quality of the data and the methodology used to collect it. Because data classification is time-consuming and labor-intensive, existing data sources from various internet repositories are commonly used. This is the case, for instance, with current large language models such as ChatGPT~\cite{openai2023chatgpt}. The credibility of these data sources is questionable, and the presence of untrustworthy data introduces the risk of adversarial attacks yielding poisoned data. In the context of CL, the adversarial attacks investigated in the current literature all aim to promote catastrophic forgetting~\cite{10657822, han2022training, pmlr-v202-kang23c, targetedForgetting}, i.e., they attack the \emph{stability} of the algorithms. Complementary test-time evasion attacks against CL models have also been reported~\cite{Bunzel_2025_b}, whereas we consider causative poisoning attacks that target future iterations.

\subsection{Motivation}
\label{sec:motivation}

The non-stationary data distribution of CL offers an attack surface for malicious input data at \emph{every} learning iteration. The concept of catastrophic forgetting considers a malicious iteration to compromise the accuracy of \emph{previous} iterations. However, models are designed with a particular domain purpose in mind, so data from subsequent iterations might already be known or collected but not yet incorporated into the training process. Consequently, an attacker aware of upcoming iterations is able to attack not only the stability, but also the \emph{plasticity} of learning new data.

To the best of our knowledge, research on attacks aiming at the plasticity of CL algorithms has not been conducted to date. This research gap needs to be investigated in order to make continual learning robust against adversarial attacks, thus ensuring the safe application of CL algorithms in critical sectors. Because these attacks are designed to hinder the learning of new knowledge, we refer to them as \textit{learning blockers}. The combination of learning blockers with the well-known phenomenon of catastrophic forgetting gives rise to a \textit{catastrophic learning} scenario.

As an illustrative example, consider a continual learning cancer detection system that must learn to identify new types of cancer as they are discovered in research. Retraining a voluminous model would prove less efficient in terms of cost and resources than training an existing model with an additional, comparatively smaller batch of new cancer data. However, using a continual learning network introduces a novel vulnerability in the form of learning blockers: the detection of novel cancers can be hindered by manipulating the data of a single iteration. If these learning blockers also result in catastrophic forgetting, the system suffers a significant deterioration in cancer detection---a catastrophic learning scenario---potentially resulting in life-threatening misclassifications within the healthcare sector.

\subsection{Contribution}
\label{sec:contribution}

This paper makes the following contributions:
\begin{itemize}
    \item We introduce the novel attack scenario of \emph{catastrophic learning}, which extends the well-known catastrophic forgetting problem by \emph{learning blockers} that attack upcoming iterations.
    \item We define a threat model tailored to continual learning, comprising the attacker's notation, goal, knowledge, and capabilities.
    \item We formulate six attacks that build learning blockers to induce catastrophic learning: two data-flipping attacks and four gradient-based attacks.
    \item We provide a detailed evaluation of these six attacks over 4{,}480 simulations on MNIST and Split-CIFAR10 against three CL algorithms.
\end{itemize}

In particular, we answer the following research questions:
\begin{enumerate}
    \item Can an adversary, by manipulating the training data of a given iteration in CL algorithms, effectively degrade the model's accuracy for that iteration or subsequent iterations?
    \item Does an attack targeting the plasticity mechanism of a CL algorithm promote catastrophic forgetting?
\end{enumerate}

\section{Background of Continual Learning}
\label{sec:background}

\subsection{Fundamentals and the Stability-Plasticity Dilemma}

A deep learning algorithm approximates a target function $f^*$ that maps an input $x$ to an output $y$ via a parameterized classifier $y = f^*(x) \approx f(x;\theta)$, where the parameter vector (weights) $\theta$ is learned. Successive applications of parameterized layers, interleaved with non-linear activation functions, build the network, and the length of this chain gives rise to the term \emph{deep} learning~\cite{deeplearning}.

Continual Learning aims to enable neural networks to approximate $f^*$ by learning from a sequence of datasets $D_1, D_2, \ldots, D_T$. The primary objective is to prevent catastrophic forgetting, which occurs when the gradient updates for a new task $t$ point in a direction antagonistic to the gradients of previous tasks. This creates the \emph{stability-plasticity dilemma}: too much stability prevents the model from learning new information (loss of plasticity), while too much plasticity leads to the erasure of old knowledge (loss of stability)~\cite{PARISI201954}.

Three major categories of strategies mitigate catastrophic forgetting:
(1) \emph{Regularization-based} strategies restrict the shift of the model's parameters in each learned iteration;
(2) \emph{Replay-based} strategies import data of old iterations to hinder parameter drift, where the replayed data may be original, synthesized, or generated by another model;
(3) \emph{Architecture-based} strategies use task-specific parameters instead of shared ones. Replay-based strategies are the most common approach~\cite{iCaRL, DBLP:journals/corr/abs-2103-16788, caccia2022new}.

\subsection{Continual Learning Scenarios}

We focus on Class-Incremental Learning (CIL), in which the model learns new classes over time and must discriminate between all seen classes during testing without the aid of task labels. Other scenarios such as Instance-Incremental, Domain-Incremental, Task-Incremental, and Online Continual Learning are discussed in the literature~\cite{wang2024comprehensive} but are out of scope for this work.

\subsection{Evaluated CL Algorithms}
\label{sec:algorithms}

We evaluate our attacks against three representative and widely used strategies:
\begin{itemize}
    \item \textbf{iCaRL}~\cite{iCaRL}: Uses a nearest-mean-of-exemplars classifier and maintains a fixed-size buffer of representative images selected via herding. Catastrophic forgetting is mitigated by computing exemplars from the current data and combining them with stored exemplars, along with a distillation loss.
    \item \textbf{DER}~\cite{DBLP:journals/corr/abs-2103-16788}: Employs a dynamically expandable representation in which a new feature extractor is concatenated to a super-feature extractor in each iteration. Previous extractors are frozen to preserve stability, while the new extractor provides plasticity.
    \item \textbf{ER-ACE}~\cite{caccia2022new}: Builds on Experience Replay and applies an asymmetric cross-entropy loss that separates the loss functions on the buffered data and the newly arrived experience, reducing the bias toward the most recent task.
\end{itemize}

\section{Threat Model and Catastrophic Learning}
\label{sec:threat}

\subsection{Definition of Catastrophic Learning}
\label{sec:cl-def}

We call our attack scenario \textit{catastrophic learning}, building on the well-known catastrophic forgetting dilemma. Building \textit{learning blockers}, we are able to create a catastrophic learning scenario for new input. We define catastrophic learning as an attack that achieves a lower training and evaluation accuracy for an attacked victim iteration. In terms of CL, this attack reduces either the plasticity or the stability tremendously, so that the rate of learned data decreases.

Catastrophic forgetting is a concept that focuses on prior iterations, whereas a learning blocker is an attack concept that focuses on compromising the current or an upcoming iteration. A learning blocker manipulates the data of one iteration in order to perturb the generalization of a CL model about itself or an upcoming iteration.

\begin{figure}[t]
    \centering
    \includegraphics[width=0.95\linewidth]{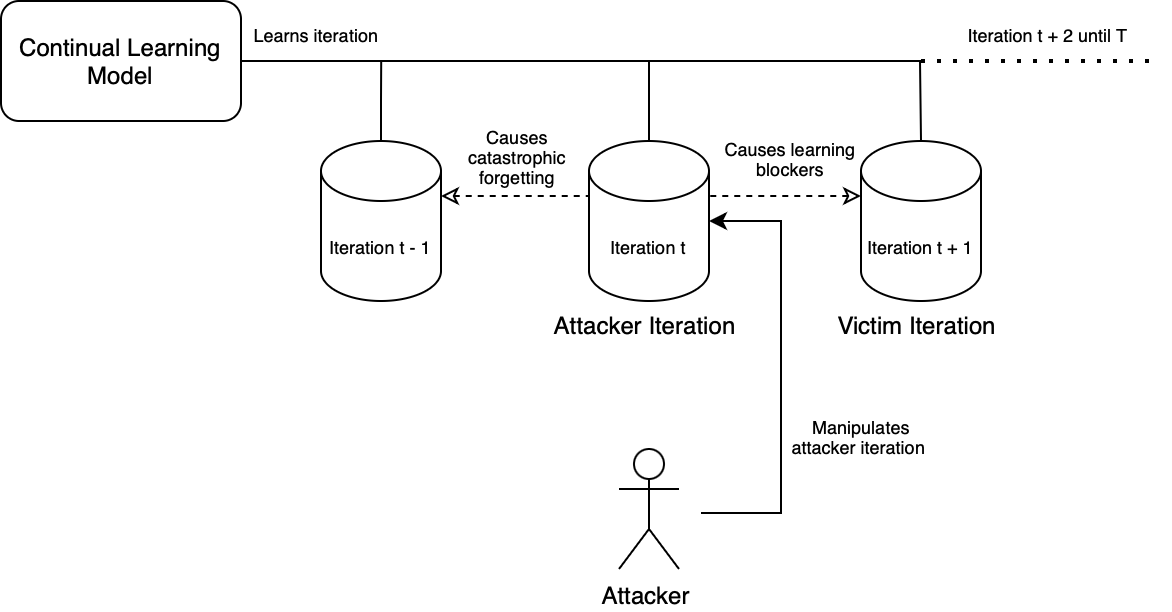}
    \caption{Exemplary representation of the catastrophic learning scenario. The attacker manipulates iteration $t$ to disrupt the upcoming victim iteration $t+1$ (plasticity) and may also induce forgetting of iteration $t-1$ (stability).}
    \label{fig:catastrophic_learning}
\end{figure}

To attack an upcoming iteration, we consider two iterations: the \textit{poisoned iteration}, from which an attack is launched, and the \textit{victim iteration}, which is corrupted. As shown in Figure~\ref{fig:catastrophic_learning}, the attacker manipulates iteration $t$ (poisoned iteration) to disrupt iteration $t+1$ (victim iteration) so that it can no longer be properly learned by the continual learning model. The construction of learning blockers is bidirectional: on the one hand, the attacker focuses on upcoming iterations, compromising the plasticity of learning new data; on the other hand, the attacker can also modify the input data to induce catastrophic forgetting of past iterations.

\subsection{Notation}
\label{sec:notation}

Based on the machine learning threat model of Demontis et al.~\cite{236234}, we formulate an extended threat model for CL. A central ingredient of that model---the transferability of adversarial perturbations across models---has since been studied and quantified for a range of architectures and domains~\cite{Bunzel_2025_e, Bunzel_2025_g, Bunzel_2025_f}, supporting the assumption that poisoned samples crafted on one model can remain effective on another. In a training phase (1), the training data is provided to the algorithm with its correct label; in a test phase (2), unseen data is given to the algorithm and its prediction is validated. CL algorithms additionally balance between plasticity (a) and stability (b). The attacker is described by her objectives (i), her knowledge of the system and its underlying algorithms (ii), and her power to manipulate the input data (iii). Data manipulation is further subdivided into having full control over the dataset (k) or only being able to manipulate some data points (l).

\subsection{Attacker's Goal}
\label{sec:goal}

Adversaries have strong incentives to manipulate CL algorithms, especially when they are deployed in critical environments. Whereas the literature typically describes an adversary's goal as the loss of prior knowledge~\cite{targetedForgetting}, we extend the adversary's goal to consider not only past but also \emph{future} iterations as vulnerable. New tasks are then learned with reduced accuracy, resulting in an unusable and unreliable model. This causes a denial-of-service condition, i.e., an \textit{availability} violation of the classification system~\cite{Barreno2010}.

\subsection{Attacker's Knowledge}
\label{sec:knowledge}

We assume that the attacker knows that the model is in a continual learning scenario. Crucial to the execution of a catastrophic learning attack is the knowledge of at least one iteration of data that is learned and at least one label of another iteration. The four PGD-based attacks additionally require the model and the CL strategy itself; however, the adversary may approximate the model using the extraction technique~\cite{10.5555/3294996.3295110}, thus moving from a black-box to a white-box scenario. Such black-box transfer is particularly relevant in practice, as transferred black-box attacks have been shown to carry non-trivial risk even when the surrogate and target models differ substantially~\cite{Bunzel_2025_g, Bunzel_2025_e}.

\subsection{Attacker's Capability}
\label{sec:capability}

To reduce the algorithm's learning capabilities, the attacker can modify training and test data. She is called \textit{causative} if she can influence the training dataset and thus the training process, and \textit{exploratory} if she can manipulate only the test data~\cite{236234}. Following Biggio et al.~\cite{poisoningattacks}, we extend the attacker capabilities: an attacker may be a malicious employee or may have gained access to the database server, giving her up to full access to the training data. Furthermore, if training is outsourced to a malicious third party, access to the model is possible without needing model extraction. If the attacker knows the purpose of the model, she can also provide manipulated information for future training---for example, adversarial patches applicable in real-world scenarios~\cite{10.1145/3665451.3665530, Bunzel_2023_c, Bunzel_face_pasting, Bunzel_2025_c} may disrupt autonomous driving systems that employ continual learning, and their effect can be amplified by environmental influences on object detection~\cite{Bunzel_2024_e}.

\section{Learning Blockers: Attack Strategies}
\label{sec:blockers}

We propose six attacks that build learning blockers for CL. Due to the balance between stability and plasticity, CL offers a new range of attack vectors beyond the often-studied catastrophic forgetting dilemma. The attacks aim to investigate this balance by injecting poisoned data in one iteration to decrease the accuracy and thus increase the loss of itself or a following iteration. Table~\ref{tab:overview} summarizes the attacks.

\begin{table}[t]
    \centering
    \scriptsize
    \caption{Overview of the proposed attacks.}
    \label{tab:overview}
    \begin{tabular}{lccc}
        \toprule
        Attack & Disruption Goal & Concept & Taxonomy \\
        \midrule
        Label-Exchange               & Stability/Plasticity & Flipping & Black-Box, Poisoning \\
        Full-Tensor-Exchange         & Stability/Plasticity & Flipping & Black-Box, Poisoning \\
        Attraction-Coincident        & Stability            & PGD      & White-Box, Poisoning \\
        Attraction-Preceding         & Stability            & PGD      & White-Box, Poisoning \\
        Repulsion-Coincident         & Plasticity           & PGD      & White-Box, Poisoning \\
        Repulsion-Preceding          & Plasticity           & PGD      & White-Box, Poisoning \\
        \bottomrule
    \end{tabular}
\end{table}

\subsection{Label-Exchange Attack}
\label{sec:label-exchange}

\begin{figure}[t]
    \centering
    \includegraphics[width=0.45\linewidth]{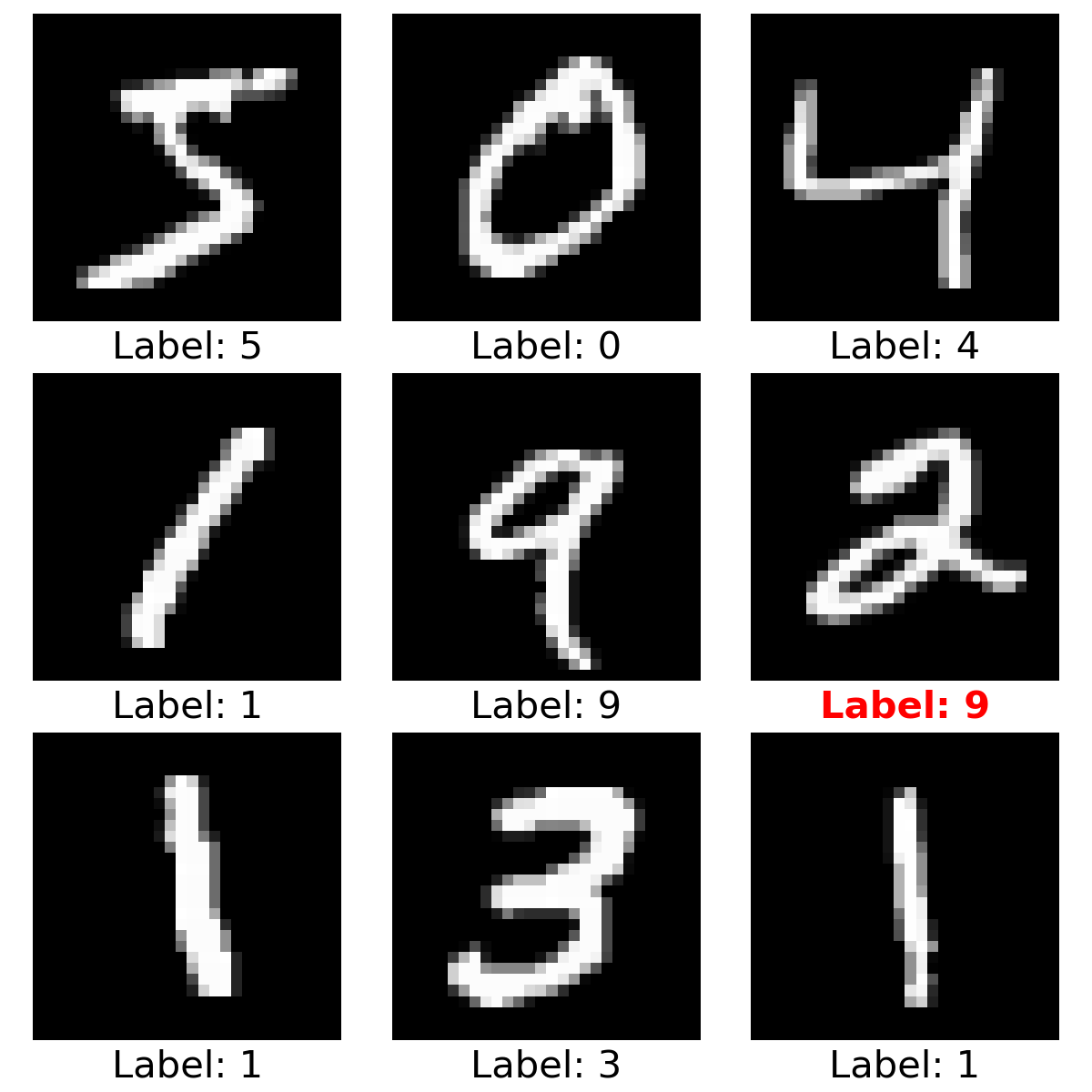}
    \caption{Exemplary MNIST data: a ``2'' labeled as ``9'' in the Label-Exchange attack.}
    \label{fig:LabelExchange}
\end{figure}

The label-exchange attack is similar to the label-flipping attack, but applied in the context of CL to cause learning blockades. The labels of an iteration controlled by the attacker (poisoned iteration $t$) are replaced by the label of another iteration (victim iteration $t+\alpha$). Hence, both iterations share the same label but contain different input data, forcing the network to learn the same label with different inputs from different iterations. Figure~\ref{fig:LabelExchange} visualizes this for the MNIST dataset: a written~``2'' is assigned the label~``9''. Because data are learned in ascending order in our test cases, the manipulated data is learned before the actual victim data with the same label.

Formally, in the context of task-incremental learning each iteration $t$ usually holds one specific label. The label-exchange attack breaks this assumption to attack plasticity:
\begin{align}\label{eq:label-exchange}
    f(x_t^i) = y_{t + \alpha} = f(x^j_{t + \alpha})
\end{align}
where $x_t^i \neq x^j_{t+\alpha}$ and, in a non-attacked scenario, $f(x_t^i) = y_t \neq y_{t+\alpha} = f(x^j_{t+\alpha})$. Due to conflicting data but coherent labels, both plasticity and stability are attacked. The underlying system and algorithm need not be known beyond the fact that we are in a CL environment.

\subsection{Full-Tensor-Exchange Attack}
\label{sec:full-tensor}

\begin{figure}[t]
    \centering
    \includegraphics[width=0.45\linewidth]{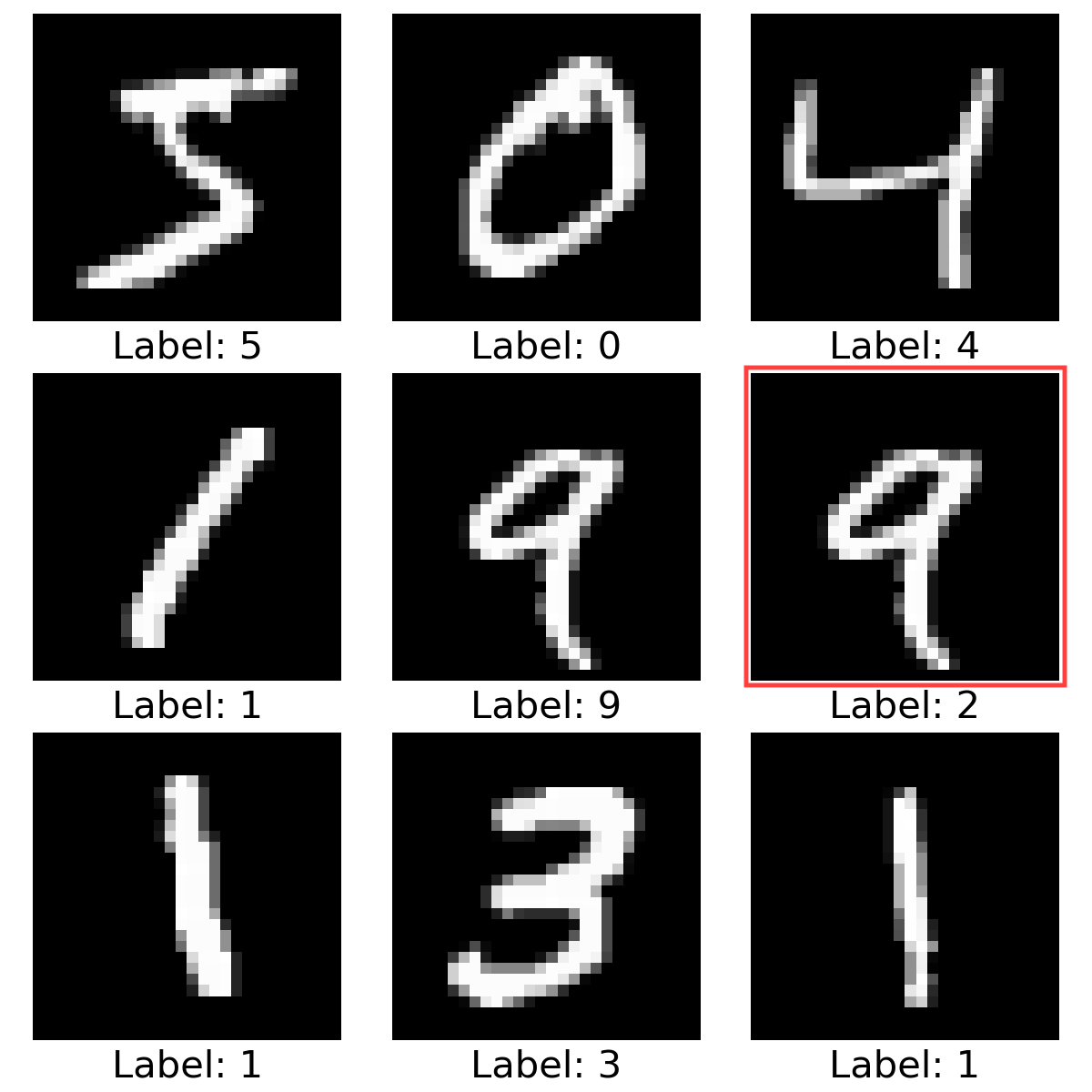}
    \caption{Exemplary MNIST data: originally a ``2'', fully exchanged by the tensor of a ``9'' in the Full-Tensor-Exchange attack.}
    \label{fig:FullTensorExchange}
\end{figure}

In the full-tensor-exchange attack, the labels of the poisoned iteration remain unmodified, but the input image of the poisoned iteration is substituted by the input image of the victim iteration. The algorithm is thus forced to learn the same input with different labels in two different iterations. Depending on the stability-plasticity dilemma, the algorithm must balance between both iterations, which results in an unpredictable classification. Figure~\ref{fig:FullTensorExchange} illustrates the manipulation on MNIST, where a written ``2'' is fully replaced by the tensor of a written ``9''.

Formally, without an attack, $x_t^i \neq x^j_{t+\alpha}$ and $f(x_t^i) \neq f(x^j_{t+\alpha})$. The attacker manipulates the data of iteration $t$ such that
\begin{align}\label{eq:full-tensor-exchange}
    f(x_t^i) = f(x^j_{t + \alpha}).
\end{align}
Both stability and plasticity are compromised, and the attacker aims to lower the accuracy for both previous and upcoming iterations without needing to know which CL algorithm is used.

\subsection{Attraction-Coincident Attack}
\label{sec:cspa}

\begin{figure}[t]
    \centering
    \includegraphics[width=0.85\linewidth]{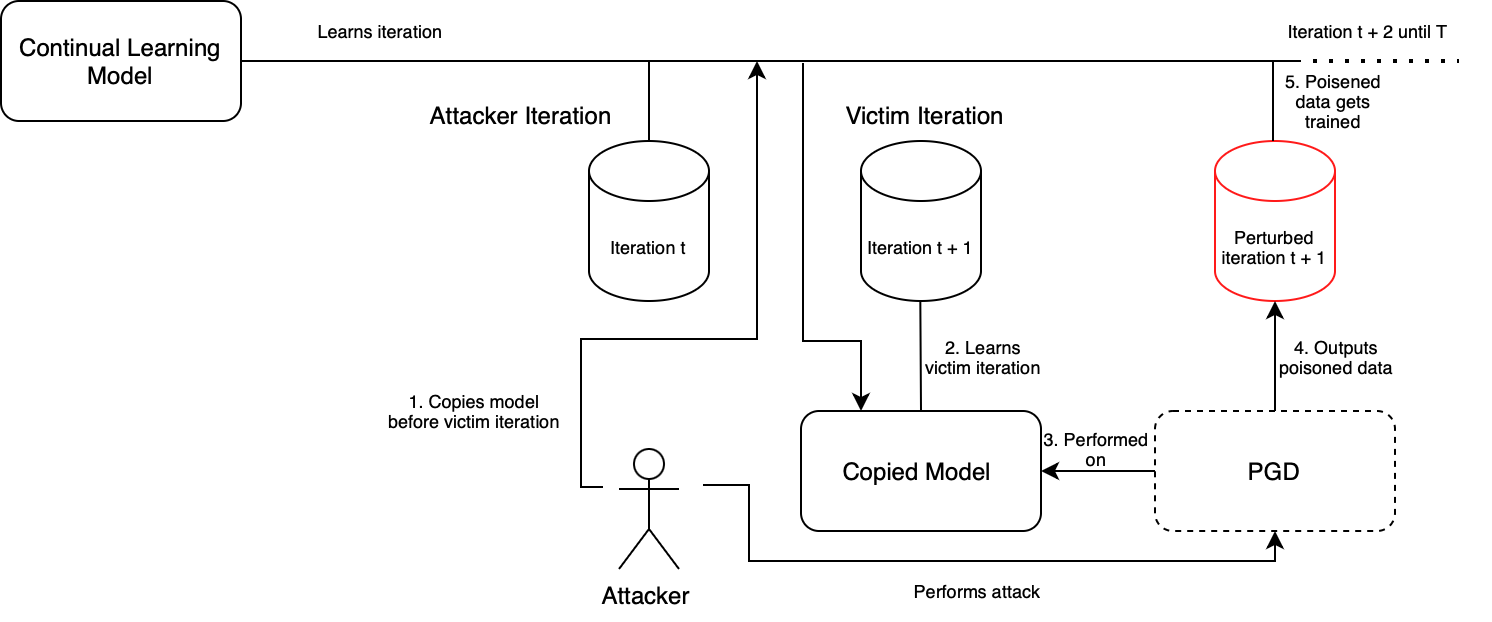}
    \caption{Coincident scenario: the surrogate is trained on only the victim iteration.}
    \label{fig:single_pretrain}
\end{figure}

The Attraction-Coincident attack brings the data of one iteration close to the label of another iteration using targeted PGD. As shown in Figure~\ref{fig:single_pretrain}, the attacker chooses iteration $t$, which is learned unaltered by the CL model. She then copies the model and trains the copy on the victim iteration $t+1$. The attacker computes the loss between the victim iteration data and the poisoned iteration's label and perturbs the victim iteration data batch using PGD so that the data batches are brought close together, choosing the perturbation with the smallest loss to the poisoned iteration's label. The original model is then trained with the adversarially manipulated data.

Bringing data batches close to the decision boundary leads to an unpredictable classification; it depends on the CL strategy whether the old or the currently learned iteration is preferred. This attacks both plasticity and stability. The adversary's input $\tilde{x}^i_{t+1}$ is computed on the target $y_t$ via
\begin{align}\label{eq:cspa}
    \tilde{x}_{t+1,n+1}^{i} = \mathrm{Proj}_{x_{t+1}+s}\!\left( \tilde{x}_{t+1,n}^{i} - \epsilon \, \frac{\nabla J_\theta(x_{t+1}^i, y_{t})}{\|\nabla J_\theta(x_{t+1}^i, y_{t})\|_2} \right)
\end{align}
\begin{align}\label{eq:compression_single}
     \tilde{x}_{t + 1}^{i} = \arg\min_{\tilde{x}_{t+1,n}^{i}}\!\left(\mathcal{L}(\tilde{x}_{t+1,n}^{i}, y_{t})\right)
\end{align}
where $\mathrm{Proj}_{x_{t+1}+s}$ denotes the projection around $x_{t+1}$ with constraint $s$ in the $L_2$-norm.

\subsection{Attraction-Preceding Attack}
\label{sec:cdpa}

\begin{figure}[t]
    \centering
    \includegraphics[width=0.85\linewidth]{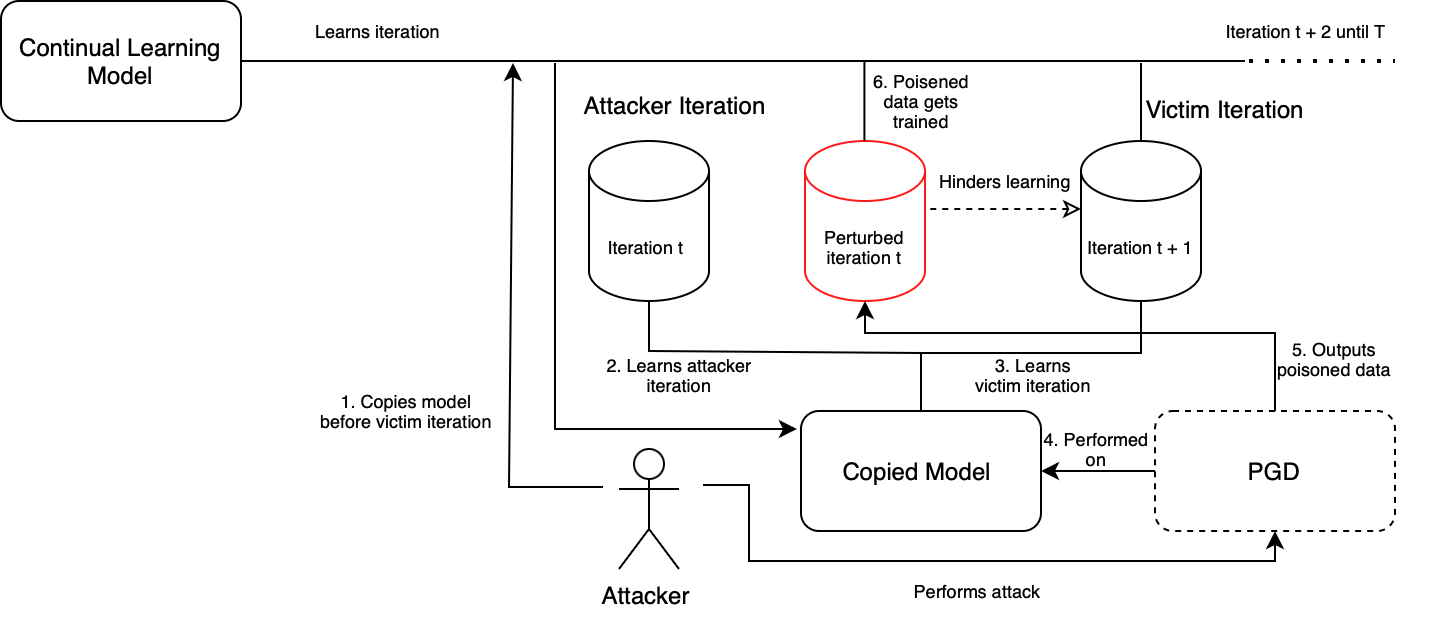}
    \caption{Preceding scenario: the surrogate is trained on both the poisoned and the victim iteration.}
    \label{fig:double_pretrain}
\end{figure}

The Attraction-Preceding attack is more resource-intensive because the attacker must pretrain two iterations on a copied model (Figure~\ref{fig:double_pretrain}). In return, the attacker is able to attack the victim iteration directly with the poisoned iteration. She copies the model and trains the copy first on the poisoned iteration $t$ and then on the victim iteration $t+1$. The calculation is the same as the targeted PGD, perturbing the poisoned iteration's data in the direction of the gradient pointing to the minimal loss to the victim iteration label. Because the produced data are pulled together in feature space, the algorithm is hindered from properly discriminating between them, lowering accuracy.

The perturbation is computed as
\begin{align}\label{eq:cdpa}
    \tilde{x}_{t,n+1}^{i} = \mathrm{Proj}_{x_{t+1}+s}\!\left( \tilde{x}_{t,n}^{i} - \epsilon \, \frac{\nabla J_\theta(x_{t}^i, y_{t+1})}{\|\nabla J_\theta(x_{t}^i, y_{t+1})\|_2} \right)
\end{align}
\begin{align}\label{eq:compression_double}
     \tilde{x}_{t}^{i} = \arg\min_{\tilde{x}_{t,n+1}^{i}}\!\left(\mathcal{L}(\tilde{x}_{t,n}^{i}, y_{t+1})\right).
\end{align}

\subsection{Repulsion-Coincident Attack}
\label{sec:hspa}

The Repulsion-Coincident attack uses only one training phase on the copied model, reducing both computation and required knowledge. Like the Attraction-Coincident attack it attempts to build a victim iteration that prevents itself from being learned, but it attacks the \emph{plasticity} of the CL strategy rather than stability. The attack pushes the victim iteration data away from the poisoned iteration's label: when the CL algorithm is trained on the malicious data, it is forced to shift its parameters to learn the input. However, CL algorithms have a built-in stability mechanism to prevent such shifts, so the new data is prevented from being learned. The computation is identical to~\eqref{eq:cspa}, but the attacker chooses the perturbation with the \emph{maximum} loss to the targeted label:
\begin{align}\label{eq:hyperextension_single}
     \tilde{x}_{t + 1}^{i} = \arg\max_{\tilde{x}_{t+1,n}^{i}}\!\left(\mathcal{L}(\tilde{x}_{t+1,n}^{i}, y_{t})\right).
\end{align}

\subsection{Repulsion-Preceding Attack}
\label{sec:hdpa}

The Repulsion-Preceding attack applies PGD in our scenario but, opposite to the Attraction variant, maximizes the loss between the poisoned iteration's data and the victim iteration's label. A copy of the current model is trained with the poisoned iteration $t$ and then the victim iteration $t+1$. The loss between the poisoned iteration's data and the victim iteration label is maximized using the iterative PGD with the $L_2$-norm; the newly generated data with maximized loss is trained into the original model. Due to the separation of the data in feature space, the attack aims to break the possible plasticity of the algorithm:
\begin{align}\label{eq:hyperextension_double}
     \tilde{x}_{t}^{i} = \arg\max_{\tilde{x}_{t,n+1}^{i}}\!\left(\mathcal{L}(\tilde{x}_{t,n}^{i}, y_{t+1})\right).
\end{align}
The perturbation step follows~\eqref{eq:cdpa}.

\section{Methodology}
\label{sec:methodology}

\subsection{Libraries}

\textbf{Avalanche}~\cite{JMLR:v24:23-0130} is an open-source end-to-end continual learning library based on PyTorch. It consists of five main modules: \textit{Benchmark} (datasets, splits, helpers), \textit{Training} (CL strategies, losses, plugins), \textit{Evaluation} (accuracy, loss, forgetting, CPU/GPU metrics), \textit{Model} (models and pretrained models), and \textit{Logging}.

\textbf{Adversarial Robustness Toolbox (ART)}~\cite{DBLP:journals/corr/abs-1807-01069} is an open-source Python library designed to evaluate, defend, and certify the robustness of machine learning models against adversarial attacks, and to facilitate the creation of adversarial examples. ART supports TensorFlow, PyTorch, Keras, and scikit-learn, which makes it compatible with Avalanche. Among its attack methods, ART includes PGD.

\subsection{Datasets}

We use two well-established benchmarks. The \textbf{MNIST} dataset~\cite{deng2012mnist} consists of $60{,}000$ training and $10{,}000$ test grayscale images of $28\times28$ pixels of handwritten digits (0--9). It is a simple dataset to develop core ideas. The \textbf{CIFAR-10} dataset~\cite{CIFAR-10} is a subset of 80~million tiny images, containing $60{,}000$ color images of $32\times32$ pixels in 10 balanced categories ($5{,}000$ training and $1{,}000$ test images per category). It is more complex than MNIST as images contain background information.

\begin{figure}[t]
  \centering
  \subfloat[MNIST]{\includegraphics[width=0.49\linewidth]{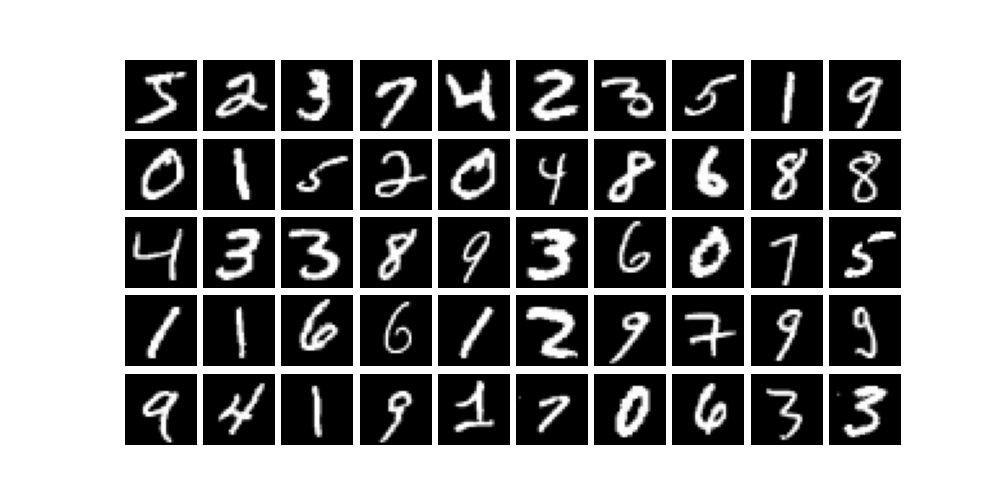}\label{fig:mnist}}
  \hfill
  \subfloat[CIFAR-10]{\includegraphics[width=0.49\linewidth]{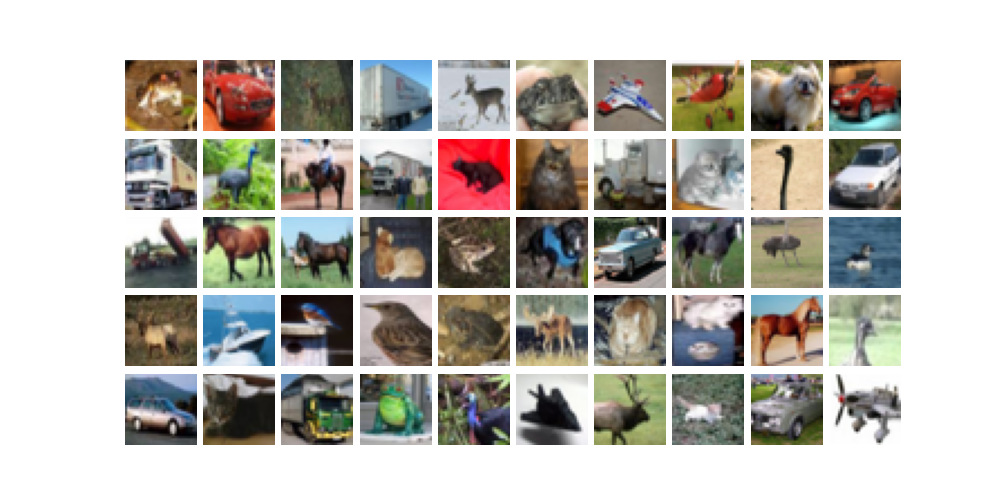}\label{fig:cifar10}}
  \caption{Examples of the benchmark datasets used in our evaluation.}
  \label{fig:datasets}
\end{figure}

\subsection{Hardware}

The MNIST simulations used an Nvidia GeForce RTX 2080 Ti with 11~GB of GPU memory and an Intel i7-8700K CPU. The CIFAR-10 simulations used an NVIDIA H100 GPU with 80~GB, 160~cores, and 2~TB of RAM, and an Intel Xeon Platinum 8380 CPU (up to 8~cores).

\subsection{Experimental Setup}
\label{sec:setup}

To systematically evaluate the attacks, we conduct separate simulations with and without the attack mode activated, starting $10$ independent simulations each and aggregating results by arithmetic mean. Since Avalanche builds on PyTorch, it is non-deterministic; we exploit this non-determinism to run $10$ different simulations. A fixed seed is used in Avalanche only so that tasks remain sorted in ascending order by label.

For both datasets, $10$ classes (tasks) are used, with the number of CL iterations set to 10 (\texttt{n\_experiences}). As a basic model, the CL algorithms DER and ER-ACE are based on a reduced implementation of \textit{ResNet18}~\cite{10.5555/3295222.3295393, facebookresearch2025gradientepisodicmemory}. The exception is iCaRL, which uses its own \textit{IcarlNet}~\cite{iCaRL} implemented in Avalanche. Both the cross-entropy loss and a copy of the model are provided to ART as a proxy classifier. The PGD attack uses the $L_2$-norm. The key parameters are listed in Table~\ref{tab:avalanche_simulation_parameter}.

\begin{table}[t]
    \centering
    \scriptsize
    \caption{Parameters provided to Avalanche and ART.}
    \label{tab:avalanche_simulation_parameter}
    \begin{tabular}{l l l}
        \toprule
        Parameter & Value & Library \\ \midrule
        shuffle & False & Avalanche\\
        seed & 0 & Avalanche\\
        n\_experiences & 10 & Avalanche \\
        criterion & Cross-entropy loss & Avalanche + ART\\
        model & Reduced ResNet18 / IcarlNet & Avalanche + ART\\
        norm & $L_2$ & ART\\
        eps & 6 & ART\\
        eps\_step & 0.3 & ART\\
        max\_iter & 40 & ART\\
        num\_random\_init & 15 & ART\\
        targeted & True & ART\\
        \bottomrule
    \end{tabular}
\end{table}

\subsection{Attack Scenarios}
\label{sec:scenarios}

We perform attacks on both MNIST and CIFAR-10. MNIST requires significantly less computation and uses only the iCaRL algorithm to reduce overhead; its purpose is to assess how far the poisoned iteration can be moved from the victim iteration while still yielding a successful attack. For the label-exchange attack we start at iteration $1$ (Avalanche's iCaRL assumes exemplars from a previous iteration). For the full-tensor-exchange we start at iteration $0$ up to $8$. From each poisoned iteration $t$, we simulate every subsequent victim iteration. This results in $36$ poisoned-victim combinations for label-exchange and $45$ for full-tensor-exchange on MNIST, plus $10$ attack-mode and $10$ comparison runs each, totaling $720$ and $900$ evaluations respectively.

For CIFAR-10, the victim iteration is exactly one iteration ahead of the poisoned iteration, with poisoned iterations ranging from $1$ to $8$ for a clear comparison across attacks. Each attack is executed using three algorithms. This yields $480$ evaluations per attack. In total, $4{,}480$ runs are performed (Table~\ref{tab:overview_evalutation_scenarios}).

\begin{table}[t]
    \centering
    \scriptsize
    \caption{Overview of the evaluation scenarios.}
    \label{tab:overview_evalutation_scenarios}
    \begin{tabular}{lccc}
        \toprule
        Attack & Dataset & Algorithm & \# Runs \\
        \midrule
        \multirow{2}{*}{Label-Exchange}               & MNIST &  iCaRL & 720 \\
         & CIFAR-10 & iCaRL, DER, ER-ACE & 480 \\
        \multirow{2}{*}{Full-Tensor-Exchange}         & MNIST &  iCaRL & 900 \\
         & CIFAR-10 & iCaRL, DER, ER-ACE & 480  \\
        Attraction-Coincident  & CIFAR-10 & iCaRL, DER, ER-ACE & 480\\
        Attraction-Preceding  &  CIFAR-10&  iCaRL, DER, ER-ACE & 480\\
        Repulsion-Coincident & CIFAR-10 & iCaRL, DER, ER-ACE &480 \\
        Repulsion-Preceding & CIFAR-10 &  iCaRL, DER, ER-ACE & 480\\
        \bottomrule
        &&& $\sum = 4{,}480$
    \end{tabular}
\end{table}

\subsection{Metrics}
\label{sec:metrics}

We use the metrics provided by Avalanche's evaluation module:
\begin{itemize}
    \item \textit{Top-1 Accuracy}: the arithmetic mean over an indicator function comparing the predicted and the true label, returning only the highest prediction. ``Accuracy'' in this paper refers to this metric.
    \item \textit{Experience Forgetting}: the difference between the accuracy obtained after the first training on a task and the accuracy after subsequent training on a different task.
\end{itemize}

\section{Evaluation}
\label{sec:evaluation}

We evaluate the six attacks. For each attack we compare the accuracy/forgetting of the attacked scenario (red) with a non-attacked comparison scenario (blue). On MNIST we examine all poisoned-victim combinations for the two flipping attacks; on CIFAR-10 we standardize to the immediate next iteration to enable a consistent comparison across all attack types, as the PGD-based attacks are limited to single-step predictions.

\subsection{Evaluation on MNIST}

\subsubsection{Label-Exchange Attack}
\begin{figure}[t]
    \centering
    \includegraphics[width=0.5\linewidth]{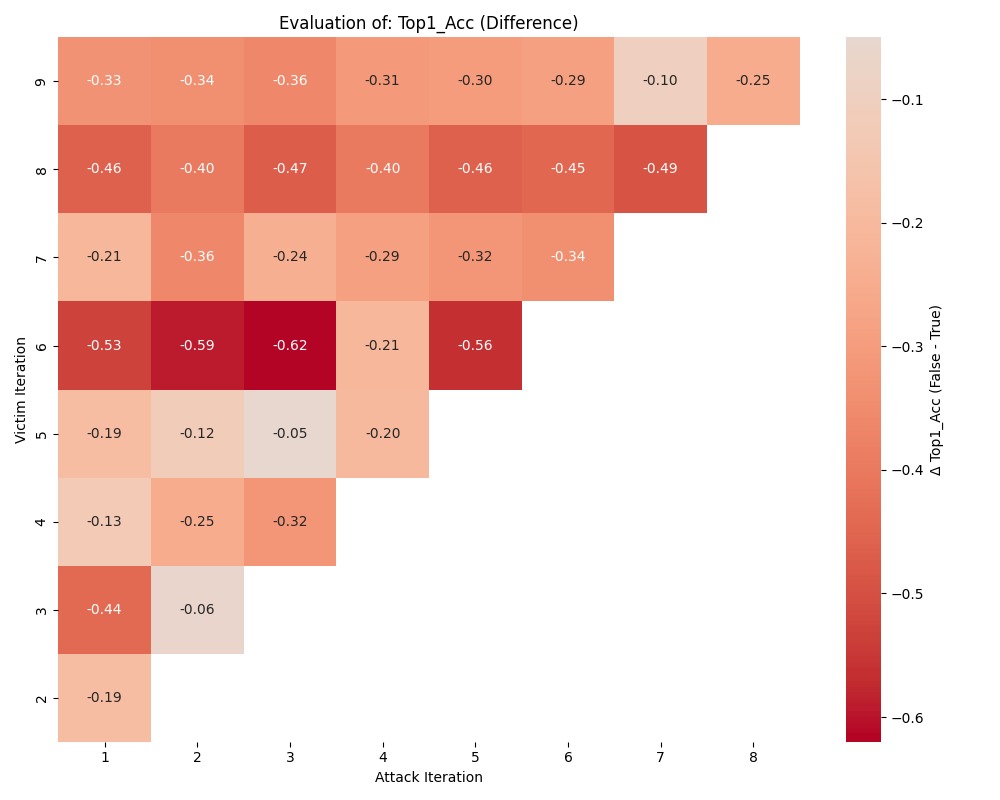}
    \caption{Difference between averaged top-1 accuracy of the victim iteration evaluation data of the attacked and unattacked scenario for the Label-Exchange attack (MNIST).}
    \label{fig:diff_top1Acc_victim_label_change}
\end{figure}

The label-exchange attack achieves a decreased accuracy on the evaluation set of the victim iteration for every pair of iterations. The resulting difference between the attacked and the unattacked scenario is shown in Figure~\ref{fig:diff_top1Acc_victim_label_change}; the attack results in a maximum decrease in accuracy of $62\%$. No discernible pattern emerged that favored a specific distance between iterations; instead, the susceptibility varies among datasets. By modifying the poisoned iteration, the accuracy of the victim iteration can be compromised, demonstrating that plasticity can be diminished for upcoming iterations.

\subsubsection{Full-Tensor-Exchange Attack}
\begin{figure}[t]
    \centering
    \includegraphics[width=0.5\linewidth]{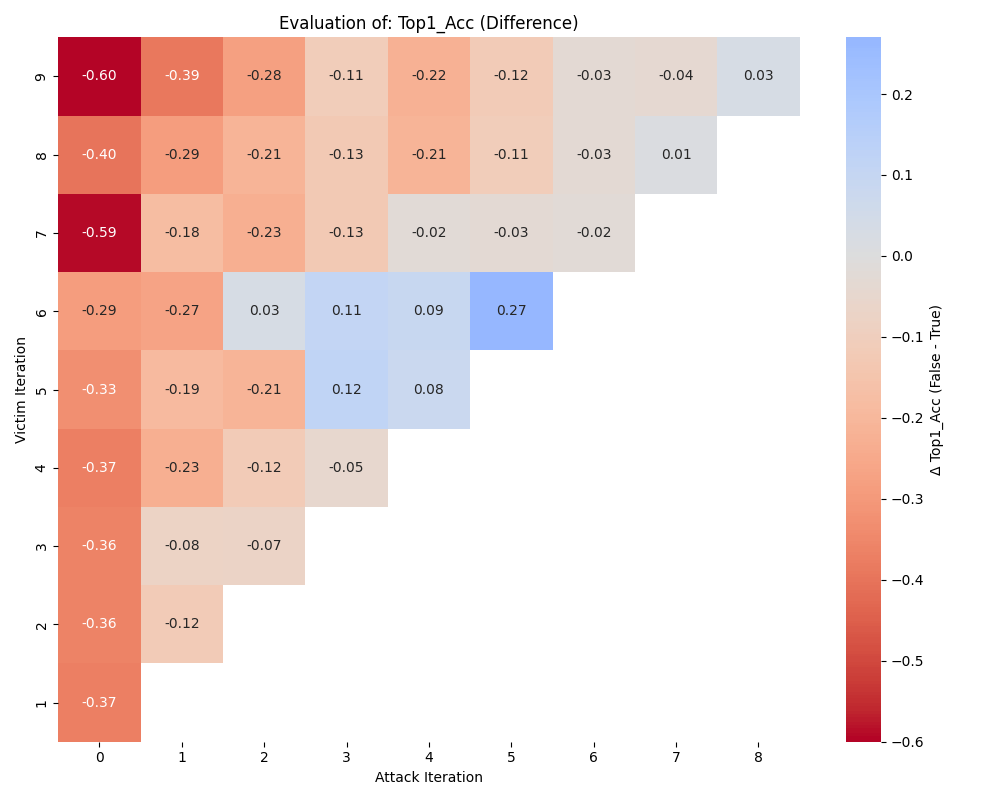}
    \caption{Difference between averaged top-1 accuracy of the victim iteration evaluation data of the attacked and unattacked scenario for the Full-Tensor-Exchange attack (MNIST).}
    \label{fig:diff_top1Acc_victim_full_tensor}
\end{figure}

iCaRL keeps a high accuracy between $61\%$ and $100\%$ in all $450$ runs in the unattacked scenario, but a huge decrease is observed when the first iteration contains the data of a subsequent iteration. The largest accuracy drop occurs when the poisoned iteration is the first iteration, and the attack is most effective when poisoned and victim iterations are far apart. The combination of poisoned iteration $0$ and victim iteration $9$ yields a total drop of $60\%$ (Figure~\ref{fig:diff_top1Acc_victim_full_tensor}). The closer the iterations, the less effective the attack, as seen on the right diagonal of the figure. The timing of the attack is thus crucial: early poisoned iterations yield high success probability.

\subsection{Evaluation on CIFAR-10}

\subsubsection{Label-Exchange Attack}
\begin{figure}[t]
  \centering
  \subfloat[DER]{\includegraphics[width=0.32\linewidth]{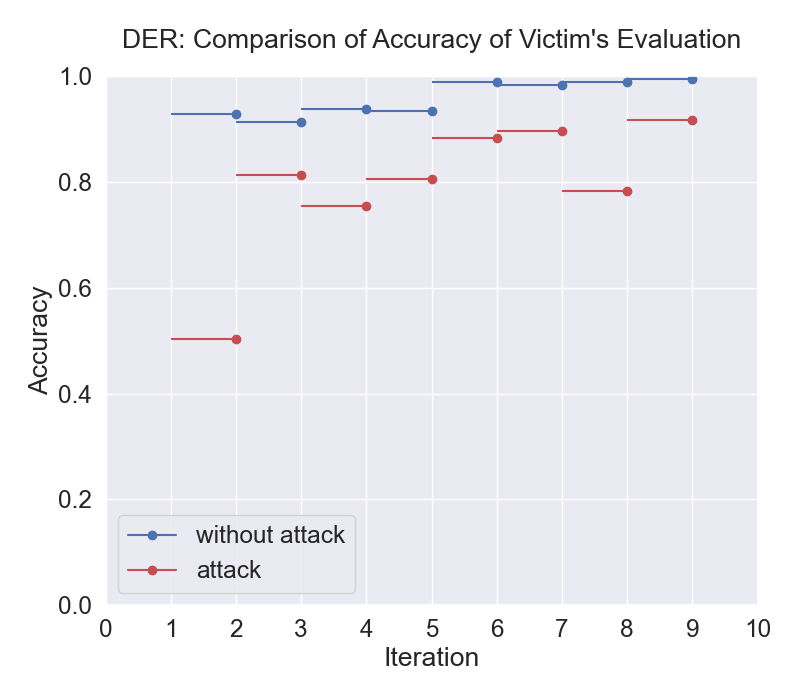}\label{fig:label_change_der}}
  \hfill
  \subfloat[ER-ACE]{\includegraphics[width=0.32\linewidth]{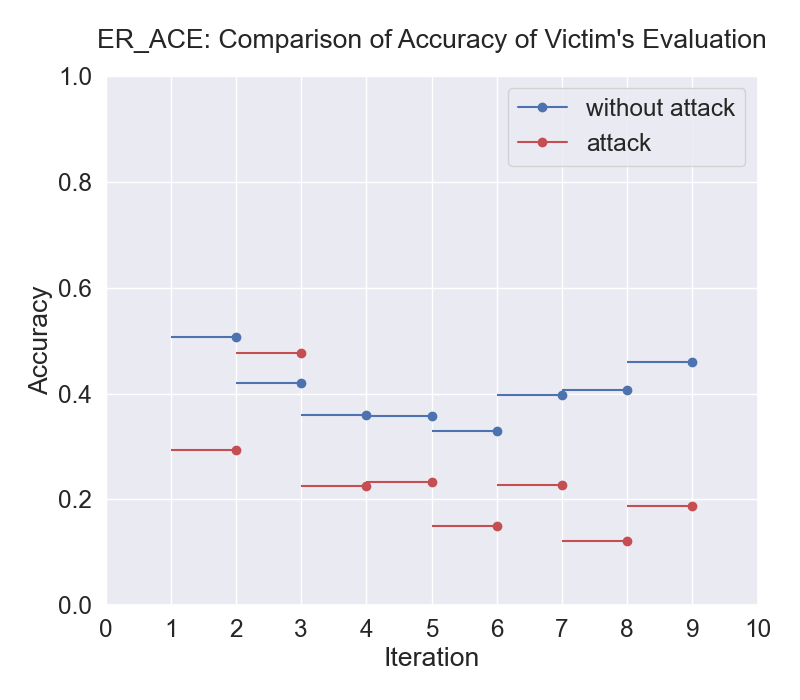}\label{fig:label_change_erace}}
  \hfill
  \subfloat[iCaRL]{\includegraphics[width=0.32\linewidth]{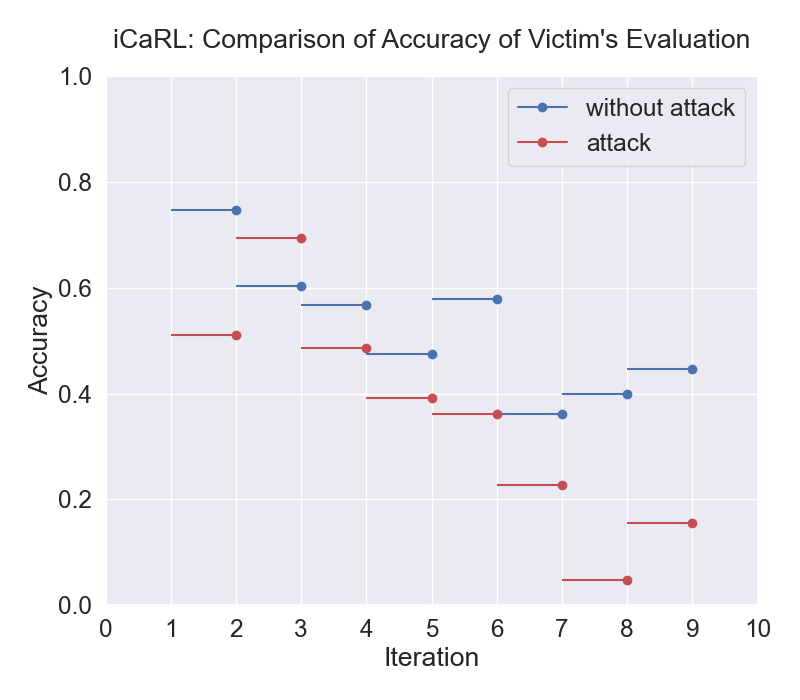}\label{fig:label_change_icarl}}
  \caption{Averaged top-1 accuracy of the victim iteration evaluation data in the Label-Exchange attack after the victim iteration (CIFAR-10).}
  \label{fig:label_change_cifar}
\end{figure}

The label-exchange attack reduces the accuracy of the victim iteration evaluation data for all considered algorithms. This is particularly evident for DER, where the attack reduces accuracy by up to $40\%$ (Figure~\ref{fig:label_change_cifar}). ER-ACE and iCaRL also show a strong decrease, except for one label. The attack is successful in reducing the plasticity of all algorithms, confirming that it is an effective technique for building learning blockers to produce a catastrophic learning scenario.

\subsubsection{Full-Tensor-Exchange Attack}
\begin{figure}[t]
  \centering
  \subfloat[DER]{\includegraphics[width=0.32\linewidth]{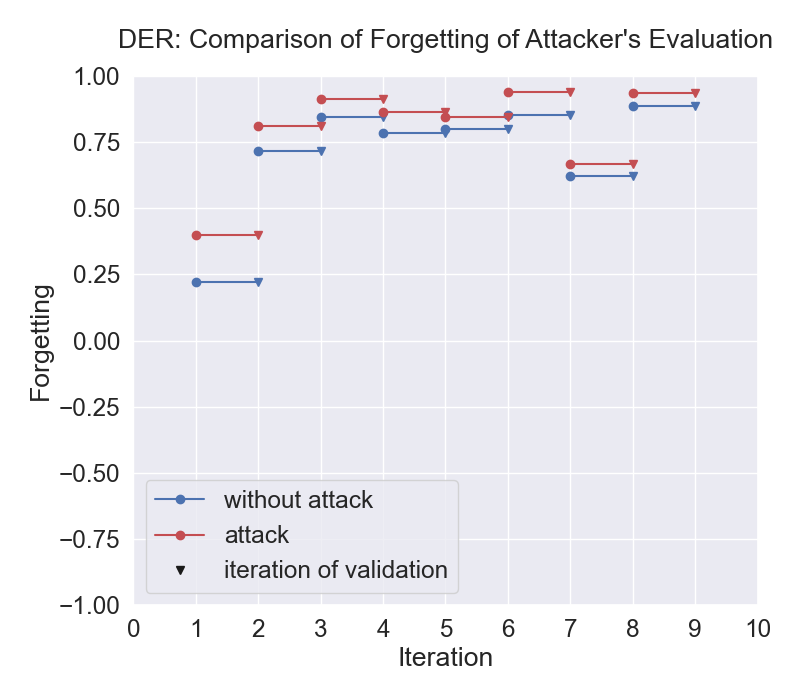}\label{fig:full_tensor_der_f}}
  \hfill
  \subfloat[ER-ACE]{\includegraphics[width=0.32\linewidth]{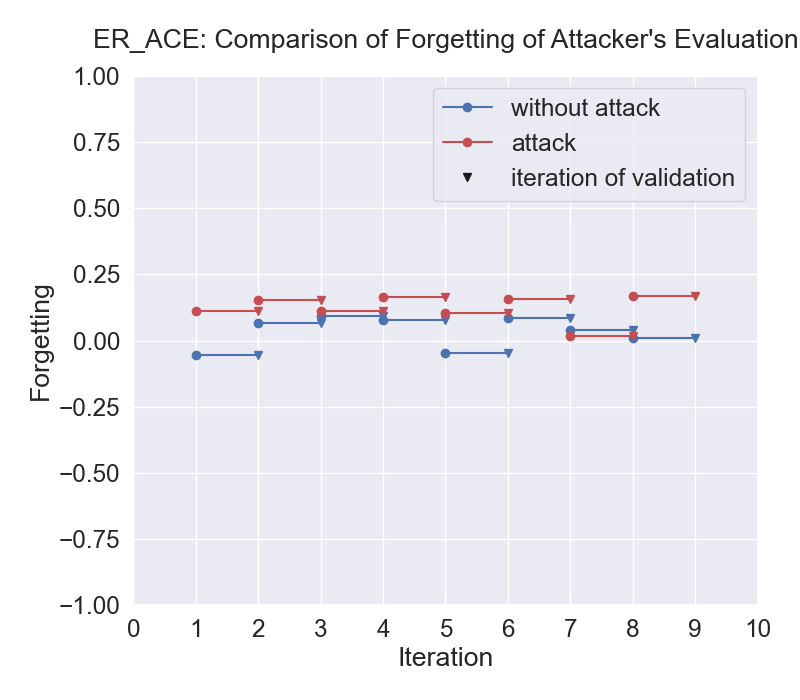}\label{fig:full_tensor_erace_f}}
  \hfill
  \subfloat[iCaRL]{\includegraphics[width=0.32\linewidth]{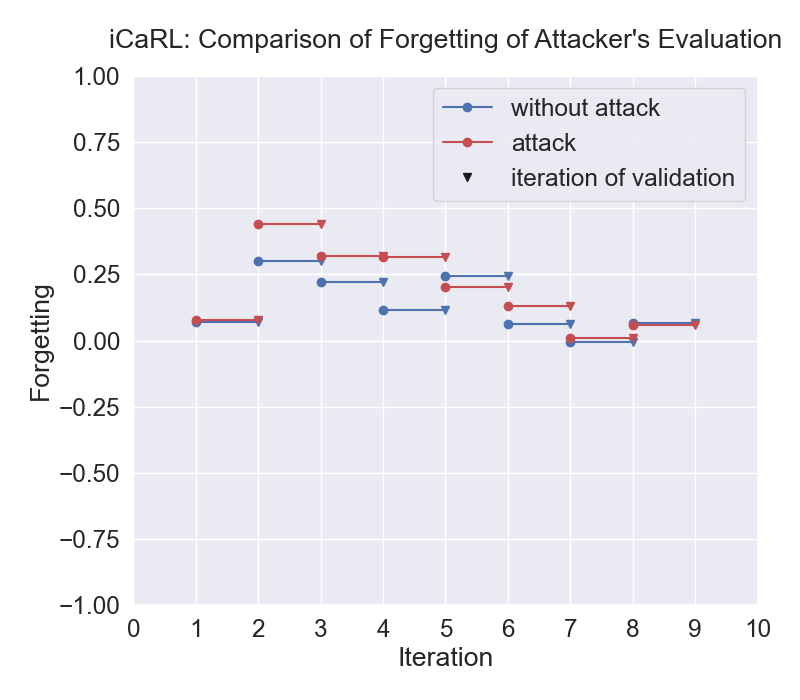}\label{fig:full_tensor_icarl_f}}
  \caption{Averaged forgetting of the poisoned iteration's evaluation data in the Full-Tensor-Exchange attack after the victim iteration (CIFAR-10).}
  \label{fig:full_tensor_forgetting}
\end{figure}

While the accuracy of the victim iteration evaluation data increases for the first iteration of DER and remains close to the same for subsequent iterations, both replay-based algorithms exhibit a partial decrease in accuracy. Crucially, forgetting could be significantly increased by the attack for all three algorithms (Figure~\ref{fig:full_tensor_forgetting}), reaching up to approximately $25\%$. The stability of the model is thus affected after the victim iteration. In summary, both plasticity and stability can be attacked; the strength in terms of plasticity depends on the algorithm, while stability can be attacked with varying strength for all algorithms.

\begin{figure}[t]
  \centering
  \subfloat[DER]{\includegraphics[width=0.32\linewidth]{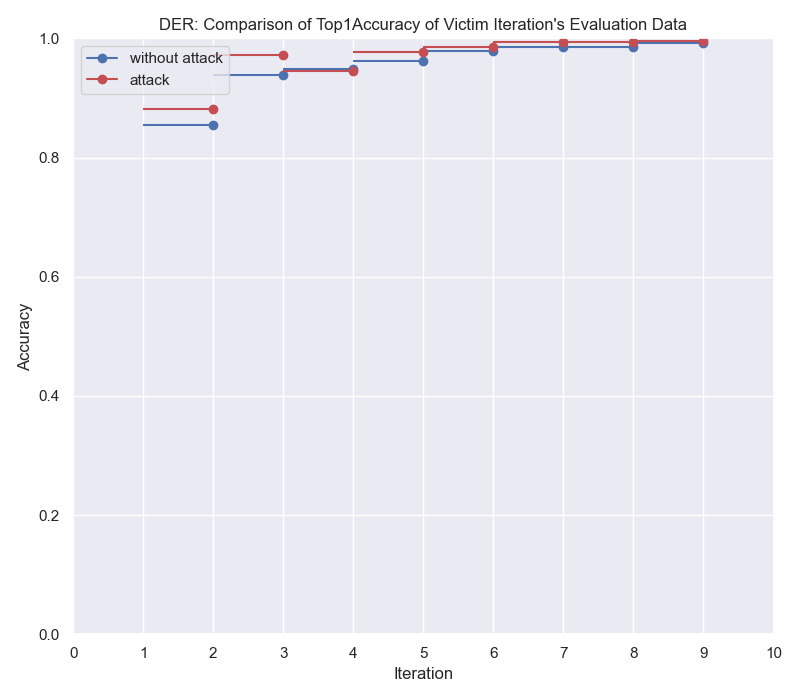}\label{fig:full_tensor_der_v}}
  \hfill
  \subfloat[ER-ACE]{\includegraphics[width=0.32\linewidth]{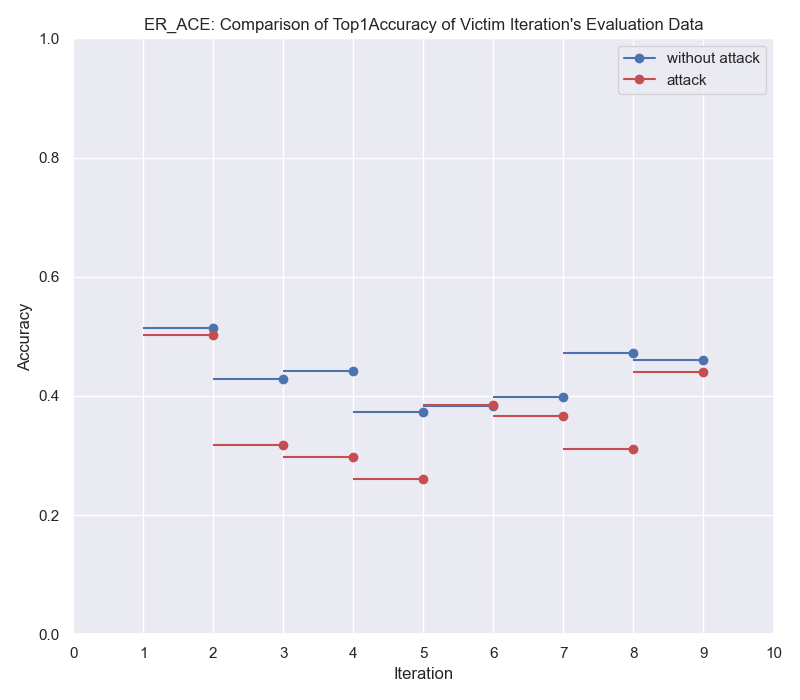}\label{fig:full_tensor_erace_v}}
  \hfill
  \subfloat[iCaRL]{\includegraphics[width=0.32\linewidth]{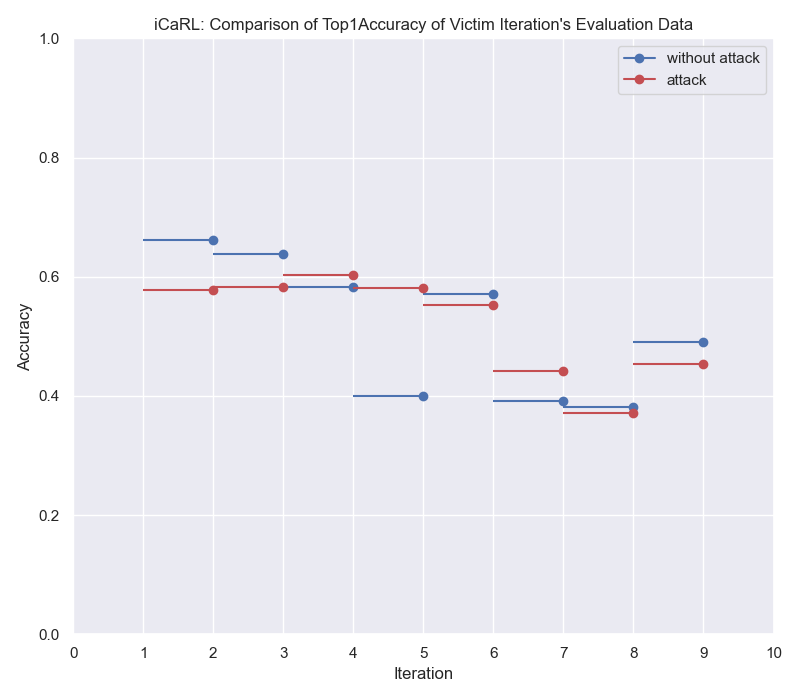}\label{fig:full_tensor_icarl_v}}
  \caption{Averaged top-1 accuracy of the victim iteration evaluation data in the Full-Tensor-Exchange attack after the victim iteration (CIFAR-10).}
  \label{fig:full_tensor_victim}
\end{figure}

\subsubsection{Attraction-Coincident Attack}
\begin{figure}[t]
  \centering
  \subfloat[DER]{\includegraphics[width=0.32\linewidth]{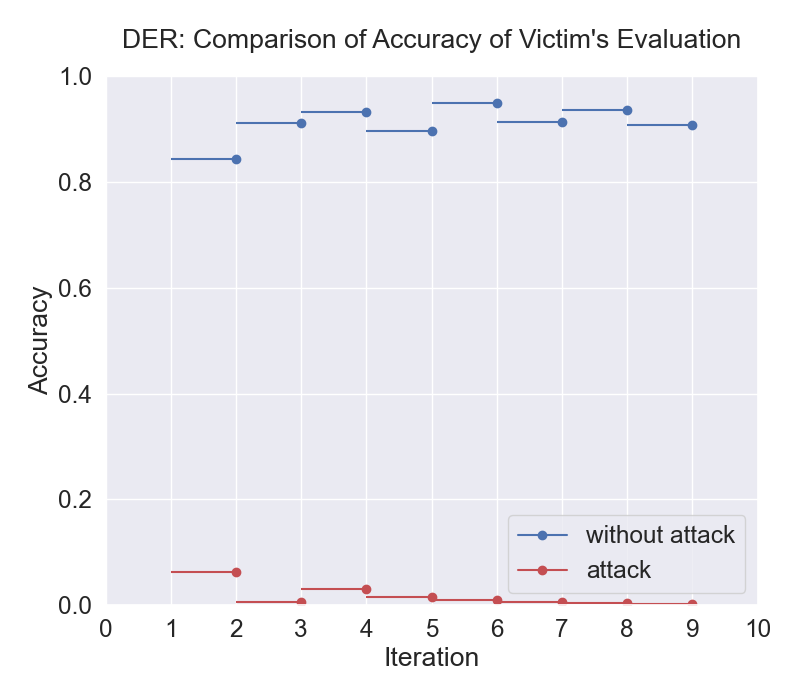}\label{fig:comp_single_der}}
  \hfill
  \subfloat[ER-ACE]{\includegraphics[width=0.32\linewidth]{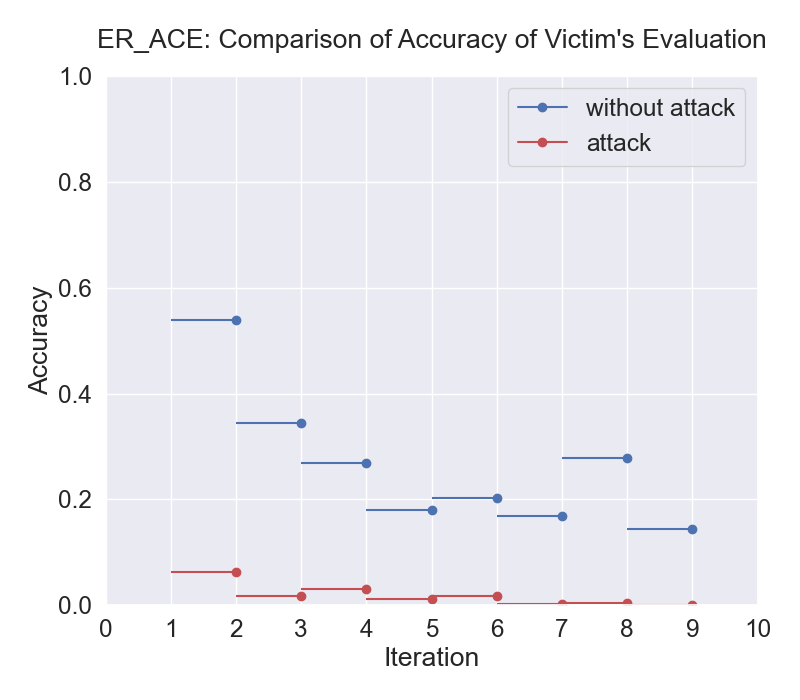}\label{fig:comp_single_erace}}
  \hfill
  \subfloat[iCaRL]{\includegraphics[width=0.32\linewidth]{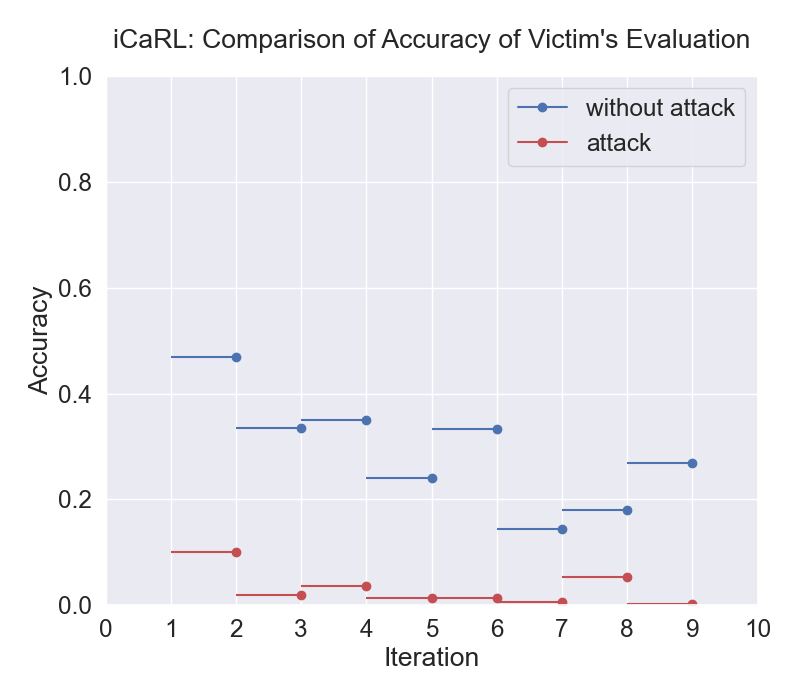}\label{fig:comp_single_icarl}}
  \caption{Averaged top-1 accuracy of the victim iteration evaluation data in the Attraction-Coincident attack after the victim iteration (CIFAR-10).}
  \label{fig:comp_single_victim}
\end{figure}

For all three algorithms, the top-1 accuracy of the victim iteration evaluation dataset is reduced across iterations. Since the training dataset of the victim iteration is slightly perturbed, the accuracy of the evaluation dataset is reduced to near zero, resulting in low adversarial robustness and reduced plasticity (Figure~\ref{fig:comp_single_victim}). The perturbations are minor and difficult for a human observer to recognize, similar to the original PGD attack. The stability of DER could not be attacked, but the stability of ER-ACE and iCaRL could be deteriorated in over $60\%$ of the iterations.

\subsubsection{Attraction-Preceding Attack}
\begin{figure}[t]
  \centering
  \subfloat[DER]{\includegraphics[width=0.32\linewidth]{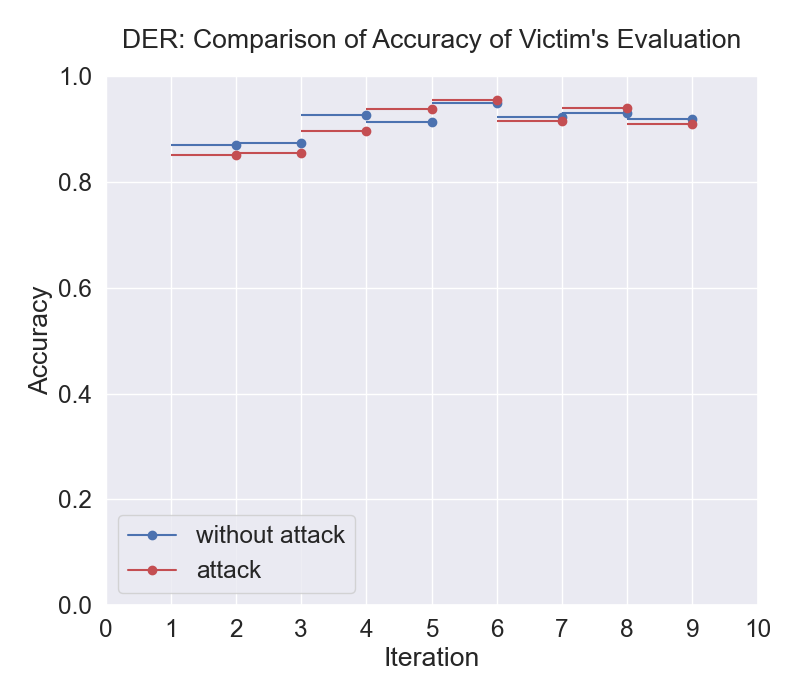}\label{fig:comp_double_der}}
  \hfill
  \subfloat[ER-ACE]{\includegraphics[width=0.32\linewidth]{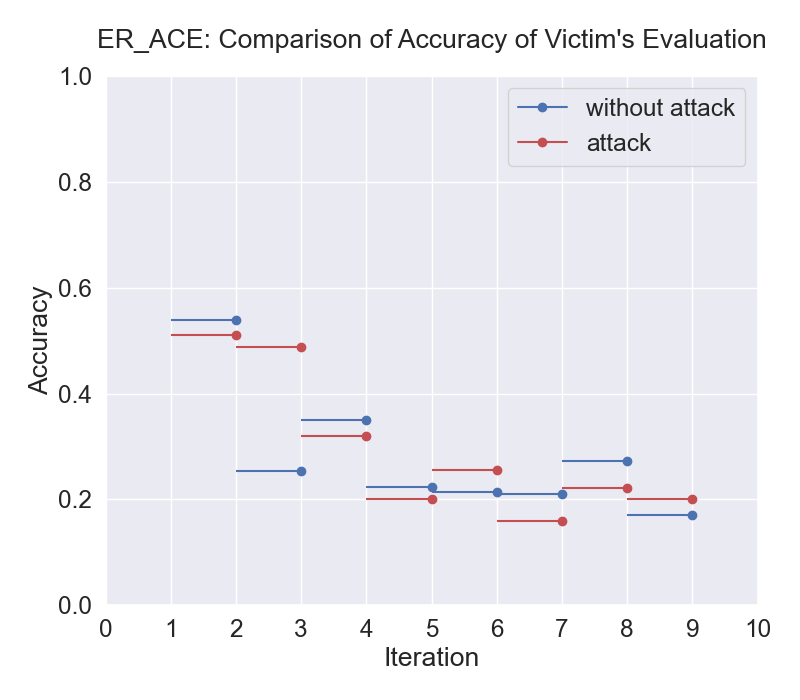}\label{fig:comp_double_erace}}
  \hfill
  \subfloat[iCaRL]{\includegraphics[width=0.32\linewidth]{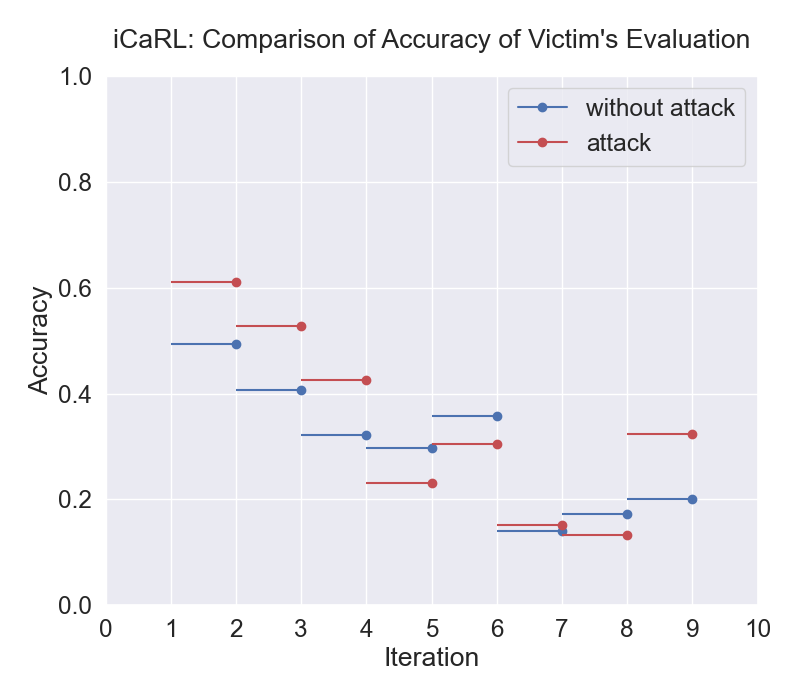}\label{fig:comp_double_icarl}}
  \caption{Averaged top-1 accuracy of the victim iteration evaluation data in the Attraction-Preceding attack after the victim iteration (CIFAR-10).}
  \label{fig:comp_double_victim}
\end{figure}

The accuracy of the victim iteration for DER aligns with the comparison scenario. For ER-ACE, the accuracy of five out of eight combinations is marginally reduced, while one combination is increased by over $20\%$. For iCaRL, five of eight combinations show a marginal to significant enhancement in accuracy, while only three result in a decrease (Figure~\ref{fig:comp_double_victim}). This supports the thesis of Bai et al.~\cite{NEURIPS2024Gradient} that the approximation of tasks can reduce forgetting. However, at the cost of reduced forgetting, a significant degradation on the evaluation dataset of the poisoned iteration takes place---corresponding to the behavior of the PGD attack.

\subsubsection{Repulsion-Coincident Attack}
\begin{figure}[t]
  \centering
  \subfloat[DER]{\includegraphics[width=0.32\linewidth]{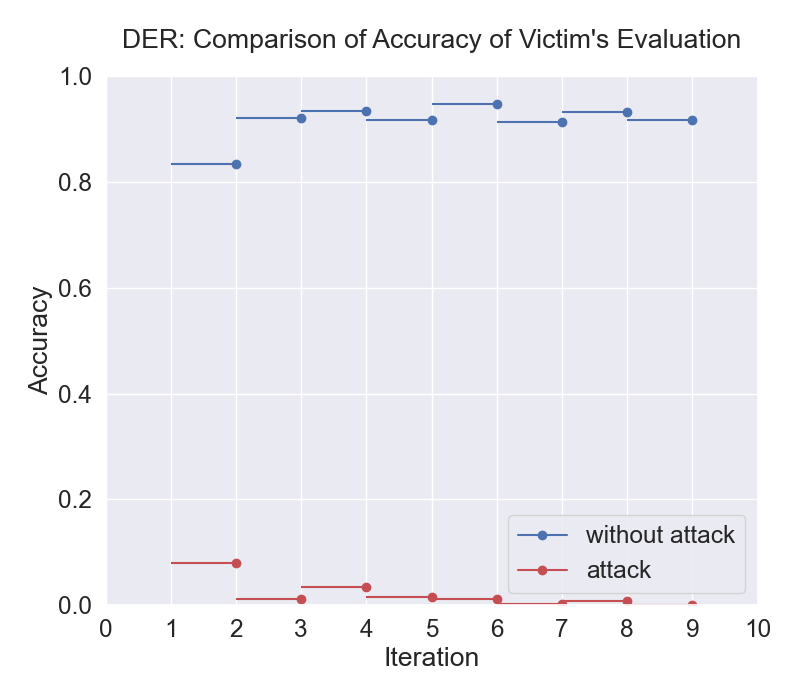}\label{fig:hyp_single_der}}
  \hfill
  \subfloat[ER-ACE]{\includegraphics[width=0.32\linewidth]{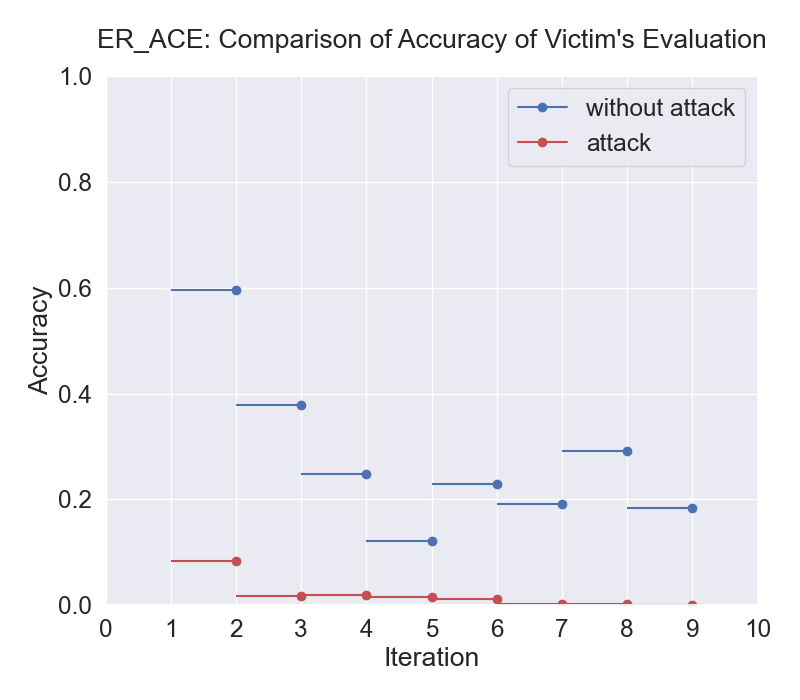}\label{fig:hyp_single_erace}}
  \hfill
  \subfloat[iCaRL]{\includegraphics[width=0.32\linewidth]{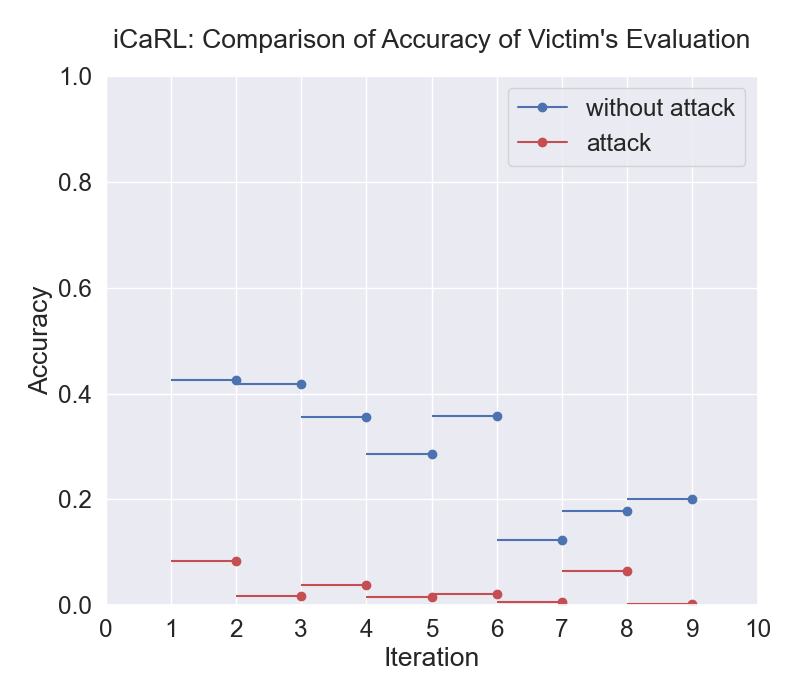}\label{fig:hyp_single_icarl}}
  \caption{Averaged top-1 accuracy of the victim iteration evaluation data in the Repulsion-Coincident attack after the victim iteration (CIFAR-10).}
  \label{fig:hyp_single_victim}
\end{figure}

All three considered algorithms demonstrate near-zero accuracy in the attacked scenario for the victim iteration evaluation data, with one exception in iteration~$2$ (Figure~\ref{fig:hyp_single_victim}). The DER algorithm performs above $80\%$ in the non-attacked scenario, the strongest contrast. By perturbing the data, the attack achieves a greatly reduced accuracy on the victim iteration evaluation data and is thus a successful poisoning attack that affects the plasticity of the CL algorithm. With regard to forgetting, the stability of some iterations could be affected, especially for iCaRL. For ER-ACE and iCaRL, maximizing the loss between poisoned and victim label reduces the accuracy of the unchanged poisoned data for most iterations---a decline in plasticity and stability that collectively contributes to a catastrophic learning scenario.

\subsubsection{Repulsion-Preceding Attack}
\begin{figure}[t]
  \centering
  \subfloat[DER]{\includegraphics[width=0.32\linewidth]{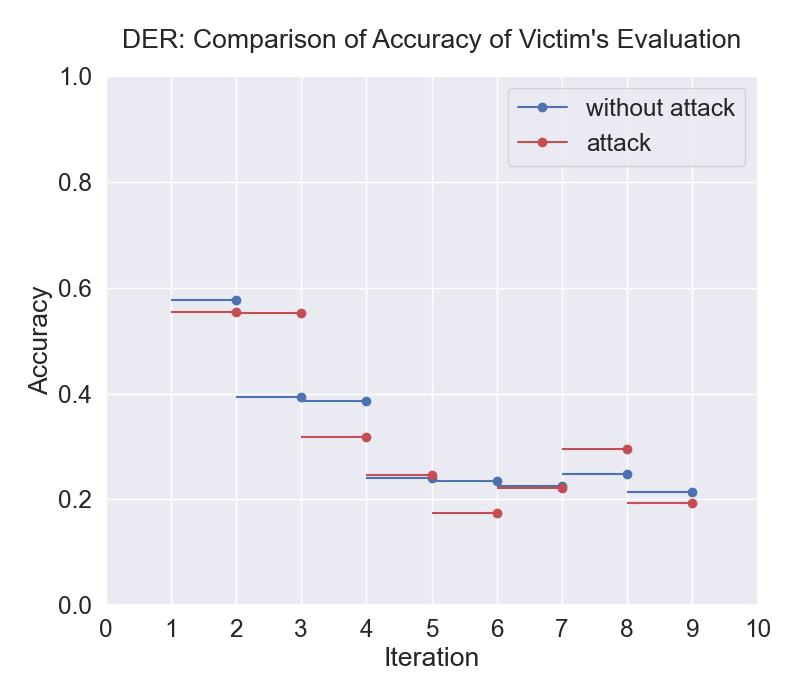}\label{fig:hyp_double_der}}
  \hfill
  \subfloat[ER-ACE]{\includegraphics[width=0.32\linewidth]{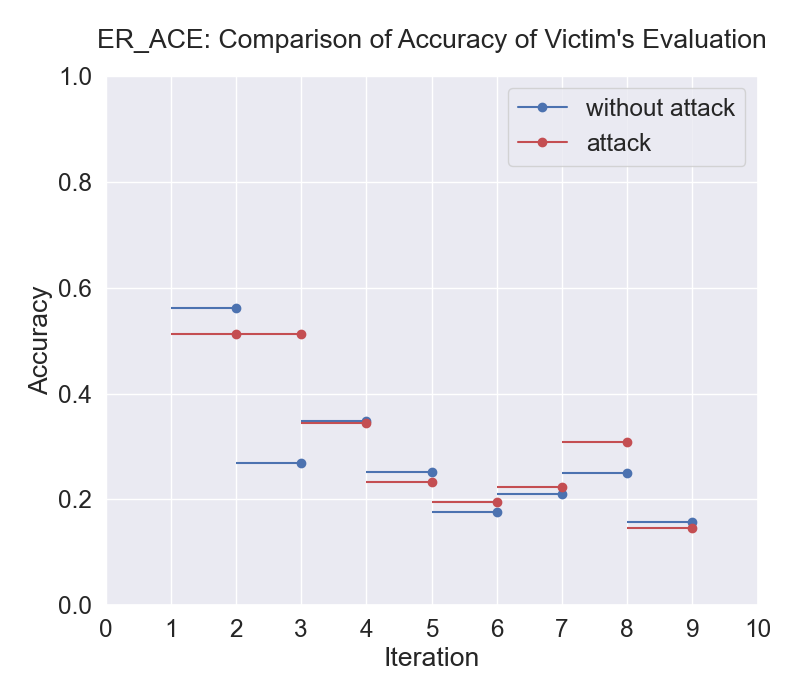}\label{fig:hyp_double_erace}}
  \hfill
  \subfloat[iCaRL]{\includegraphics[width=0.32\linewidth]{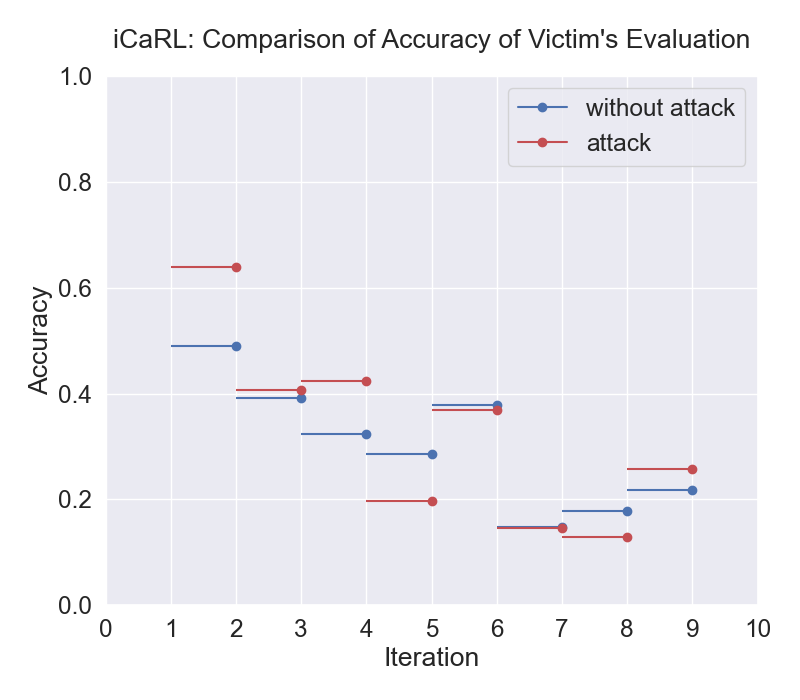}\label{fig:hyp_double_icarl}}
  \caption{Averaged top-1 accuracy of the victim iteration evaluation data in the Repulsion-Preceding attack after the victim iteration (CIFAR-10).}
  \label{fig:hyp_double_victim}
\end{figure}

For DER, four out of eight simulations result in decreased accuracy on the victim iteration evaluation dataset, with an outlier showing a strong increase (Figure~\ref{fig:hyp_double_victim}). The same outlier is observed for ER-ACE, while the remaining iterations mostly align. For iCaRL there are three outliers, two generating an increase and one a decrease. The attack simulations exhibit average forgetting values below zero for all three algorithms, indicating that the augmentation of the vector space can mitigate forgetting. Simultaneously, the attack causes a significant decrease in the accuracy of the poisoned iteration's labels, reaching up to $70\%$. Depending on the iteration, the plasticity in all algorithms can be reduced by up to $10\%$. The changes are slight and difficult for a human observer to perceive.

\subsection{Stealthiness of the PGD-based Attacks}
\label{sec:stealth}

\begin{figure}[t]
  \centering
  \subfloat[Original $x$]{\includegraphics[width=0.32\linewidth]{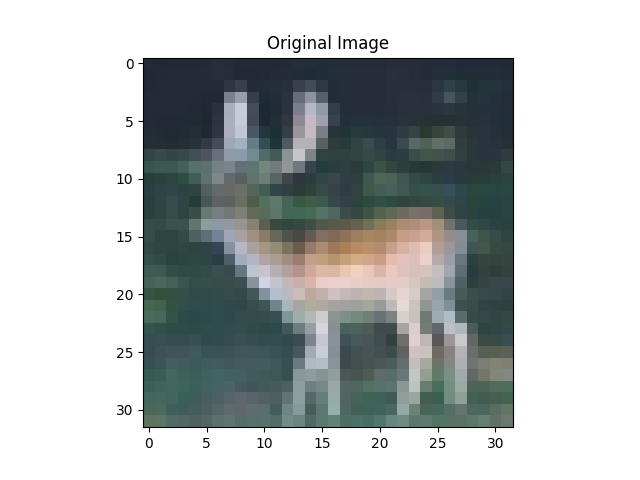}\label{fig:orignal}}
  \hfill
  \subfloat[Perturbation $\delta$]{\includegraphics[width=0.32\linewidth]{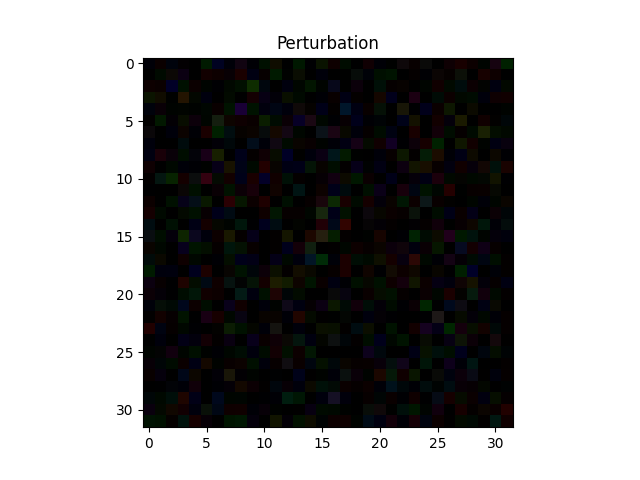}\label{fig:perturbation_noise}}
  \hfill
  \subfloat[Adversarial $\tilde{x}$]{\includegraphics[width=0.32\linewidth]{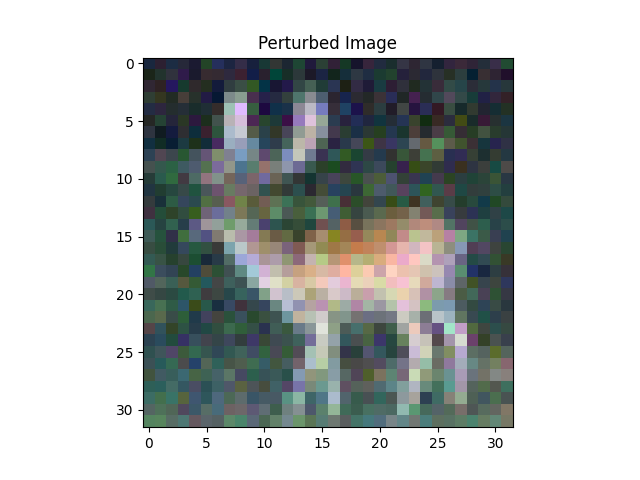}\label{fig:perturbed}}
  \caption{Perturbation produced by the Attraction-Coincident attack on CIFAR-10: the perturbation in (b) is added to the original in (a) to obtain the adversarial image in (c).}
  \label{fig:perturbation}
\end{figure}

Because the PGD-based attacks only slightly perturb the data, the changes are difficult for a human observer to perceive. On average, fewer than half of the $5{,}000$ samples per CIFAR-10 iteration are perturbed (Table~\ref{tab:num_pertrubations}), resulting in an inconspicuous attack through PGD. A variety of detection and mitigation techniques for adversarial noise and patch attacks have been proposed in the literature, including high-entropy region detection~\cite{Bunzel_2023_b}, signal-based features~\cite{Bunzel_2024_b}, depth contrast~\cite{Bunzel_2024_d}, statistical feature-based analysis~\cite{Bunzel_2025_d}, and runtime-constrained image preprocessing defenses~\cite{Bunzel_2024_g}; the broader challenge of detecting adversarial examples in real-world deployments is discussed in~\cite{Bunzel_2024_c}. Adapting such detectors to the continual-learning setting, where poisoned samples target not the current but a future iteration, is an open problem that our threat model makes explicit. Figure~\ref{fig:perturbation} exemplifies the original image, the perturbation, and the resulting adversarial image for the Attraction-Coincident attack on CIFAR-10.

\begin{table}[t]
    \centering
    \scriptsize
    \caption{Averaged number of perturbed data points per algorithm (out of 5{,}000 per CIFAR-10 iteration).}
    \label{tab:num_pertrubations}
    \begin{tabular}{lccc}
        \toprule
        Attack & DER & ER-ACE & iCaRL \\
        \midrule
        Attraction-Coincident   & 2146.45 & 2251.34 & 2324.13 \\
        Attraction-Preceding   & 2247.24 & 2317.85 & 2391.15 \\
        Repulsion-Coincident & 2205.60 & 2287.51 & 2320.86 \\
        Repulsion-Preceding & 2334.08 & 2310.52 & 2396.95 \\
        \bottomrule
    \end{tabular}
\end{table}

\subsection{Concluding Discussion}

The comprehensive evaluation highlights the vulnerability of CL algorithms. Attacks such as Label-Exchange and Full-Tensor-Exchange consistently demonstrate their capability to impair \textit{plasticity}, especially when the attack is strategically timed in early or mid-training iterations. The Full-Tensor-Exchange attack was additionally able to significantly reduce stability, suggesting that learning of new information can be effectively reduced by targeted manipulation, and that these learning blockers can induce catastrophic forgetting as a byproduct---an overall catastrophic learning scenario.

The Attraction and Repulsion variants, particularly in their coincident form, expose a dual threat to both stability and plasticity. The stability of stored knowledge is not only vulnerable to direct interference but also indirectly affected through our contributed adversarial interference. A central insight is that the success and characteristics of an attack are highly dependent on the architecture and strategy of the target algorithm. Replay-based models such as iCaRL and ER-ACE exhibited unique failure patterns, often producing forgetting or reducing the learning of new tasks when exposed to adversarial manipulations. Table~\ref{tab:attack_summary} summarizes the empirical effects.

\begin{table}[t]
    \centering
    \scriptsize
    \caption{Summary of the evaluation results.}
    \label{tab:attack_summary}
    \begin{tabular}{lccc}
        \toprule
        Attack & Dataset & Algorithm & Attacks \\
        \midrule
        Label-Exchange & MNIST & iCaRL & Plasticity \\
         & CIFAR-10 & iCaRL, DER, ER-ACE & Plasticity \\
        Full-Tensor-Exchange & MNIST & iCaRL & Stability \& Plasticity \\
         & CIFAR-10 & iCaRL, DER, ER-ACE & Stability \& Plasticity \\
        Attraction-Coincident  & CIFAR-10 & iCaRL, DER, ER-ACE & Stability \& Plasticity \\
        Attraction-Preceding  & CIFAR-10 & iCaRL, DER, ER-ACE & Stability \\
        Repulsion-Coincident & CIFAR-10 & iCaRL, DER, ER-ACE & Stability \& Plasticity \\
        Repulsion-Preceding & CIFAR-10 & iCaRL, DER, ER-ACE & Stability \& Plasticity \\
        \bottomrule
    \end{tabular}
\end{table}

\subsection{Limitations}

The Avalanche library performs sorting and checks automatically, which facilitates a first-time user's work but required several attempts to circumvent these mechanisms in order to implement the label-exchange attack. The iterative nature of CL algorithms limits parallelization, making simulations time-consuming even on advanced hardware. Full simulations on all three algorithms, including comparison runs, took between one and seven days. The extent to which the order of tasks plays a role could not be investigated due to time constraints.

\section{Conclusion and Future Work}
\label{sec:conclusion}

This paper addresses the security of continual learning. After an introduction to continual deep learning networks and related attacks, we elaborated a major research gap: the state of the art focuses exclusively on attacks that induce catastrophic forgetting, whereas we showed that not only previous but also upcoming iterations can be attacked. We constructed a novel attack scenario, \textit{catastrophic learning}, described it with an associated threat model, and defined \textit{learning blockers} as the means to attack upcoming iterations.

To investigate the effectiveness of learning blockers, we contributed six distinct attack implementations and evaluated them over 4{,}480 simulations across three CL algorithms used as benchmarks in the literature. Our findings reveal that introducing learning blockers is indeed feasible, and that these blockers also led to catastrophic forgetting in specific iterations. We can thus answer both research questions positively: an adversary can degrade the accuracy of current or subsequent iterations by manipulating a single iteration's training data, and attacks aiming at plasticity can also negatively affect stability.

\textbf{Future work.} In both flipping attacks the entire iteration is manipulated; performing stealthier attacks on small subsets should be investigated. The efficacy of label-exchange in attacking plasticity, complemented by the full-tensor-exchange's promotion of forgetting, suggests that a combination of both should be explored. Further simulations with a different task order could reveal whether the results are iteration- or data-dependent, and larger-scale simulations covering all poisoned-victim pairs for DER and ER-ACE would refine the picture. Finally, it should be investigated to what extent known unlearning strategies can help to remove malicious iterations once learning blockers have been introduced into a CL model. Relatedly, since defenses against adversarial perturbations often trade robustness for clean accuracy~\cite{Bunzel_2026_b} and since edge cases generated to probe models share the spirit of our minimal-perturbation poisoning~\cite{Bunzel_2024_a}, future defenses should be evaluated both on the accuracy cost they impose and on their coverage of the edge-case-like samples produced by learning blockers.

\bibliographystyle{IEEEtran}
\bibliography{references}

\end{document}